\documentclass[twocolumn, twocolappendix]{aastex63}

\usepackage{graphicx}	
\usepackage{color}
\usepackage{amsmath}	
\usepackage{amssymb}	
\usepackage{mathrsfs}
\usepackage{mathrsfs}
\usepackage{multirow}

\newcommand{\msun}{M_\odot}
\newcommand{\rsun}{R_\odot}

\newcommand{\mpc}{{\rm Mpc}}

\newcommand{\pc}{{\rm pc}}
\newcommand{\mum}{\mu {\rm m}}
\newcommand{\kms}{{\rm km~s}^{-1}}

\newcommand{\Muv}{M_{\rm UV}}

\newcommand{\K}{{\rm K}}
\newcommand{\beq}{\begin{equation}}
\newcommand{\eeq}{\end{equation}}

\shorttitle{High-redshift Tidal Disruption Events}
\shortauthors{Inayoshi et al.}

\begin{document}

\title{
Exploring Low-Mass Black Holes through Tidal Disruption Events in the Early Universe:\\
Perspectives in the Era of JWST, RST, and LSST Surveys
}

\correspondingauthor{Kohei Inayoshi}
\email{inayoshi@pku.edu.cn}

\author[0000-0001-9840-4959]{Kohei Inayoshi}
\affiliation{Kavli Institute for Astronomy and Astrophysics, Peking University, Beijing 100871, China}

\author[0000-0003-4299-8799]{Kazumi Kashiyama}
\affiliation{Department of Physics, Graduate School of Science, University of Tokyo, Bunkyo-ku, Tokyo 113-0033, Japan}
\affiliation{Kavli Institute for the Physics and Mathematics of the Universe (Kavli IPMU, WPI), The University of Tokyo, Chiba 277-8583, Japan}

\author[0000-0002-1044-4081]{Wenxiu Li}
\affiliation{Kavli Institute for Astronomy and Astrophysics, Peking University, Beijing 100871, China}
\affiliation{Department of Astronomy, School of Physics, Peking University, Beijing 100871, China}

\author[0000-0002-6047-430X]{Yuichi Harikane}
\affiliation{Institute for Cosmic Ray Research, The University of Tokyo, 5-1-5 Kashiwanoha, Kashiwa, Chiba 277-8582, Japan}

\author[0000-0002-4377-903X]{Kohei Ichikawa}
\affiliation{Global Center for Science and Engineering, Faculty of Science and Engineering, Waseda University}
\affiliation{
Department of Physics, School of Advanced Science and Engineering, Faculty of Science and Engineering, Waseda University, 3-4-1,
Okubo, Shinjuku, Tokyo 169-8555, Japan}

\author[0000-0003-2984-6803]{Masafusa Onoue}
\affiliation{Kavli Institute for Astronomy and Astrophysics, Peking University, Beijing 100871, China}
\affiliation{Kavli Institute for the Physics and Mathematics of the Universe (Kavli IPMU, WPI), The University of Tokyo, Chiba 277-8583, Japan}
\affiliation{Center for Data-Driven Discovery, Kavli IPMU (WPI), UTIAS, The University of Tokyo, Chiba 277-8583, Japan}

\begin{abstract}
The James Webb Space Telescope (JWST) has recently uncovered the presence of low-luminosity active galactic nuclei (AGNs) at $z=4-11$.
Spectroscopic observations have provided estimates of the nuclear black hole (BH) masses for these sources, extending the low-mass boundary down to $M_\bullet \sim 10^{6-7}~\msun$. 
Despite this breakthrough, the observed lowest mass of BHs is still $\gtrsim 1-2$ orders of magnitude heavier than the predicted mass range of their seed population,
thereby leaving the initial mass distribution of massive BHs poorly constrained.
In this paper, we focus on UV-to-optical (in rest frame) flares of stellar tidal disruption events (TDEs) embedded in low-luminosity AGNs
as a tool to explore low-mass BH populations with $\lesssim 10^{4-6}~\msun$.
We provide an estimate of the TDE rate over $z=4-11$ associated with the properties of JWST-detected AGN host galaxies, and find that 
deep and wide survey programs with JWST and Roman Space Telescope (RST) can detect and identify TDEs up to $z\simeq 4-7$.
The predicted detection numbers of TDEs at $z>4$ in one year are $\mathcal{N}_{\rm TDE} \sim 2-10~(0.2-2)$ for the JADES-Medium (and COSMOS-Web) survey with JWST, and 
$\mathcal{N}_{\rm TDE} \sim 2-10~(8-50)$ for the Deep (and Wide) tier of the High-latitude Time Domain Survey with RST.
We further discuss the survey strategies to hunt for the transient high-redshift TDEs in wide-field surveys with RST, 
as well as a joint observation campaign with the Vera C. Rubin Observatory for enhancing the detection number.
The high-redshift TDE search will give us a unique opportunity to probe the mass distribution of early BH populations.
\end{abstract}


\section{introduction}

Observations of high-redshift quasars within the first few billion years of cosmic history have been benefited from wide-field surveys 
such as the Sloan Digital Sky Survey (SDSS; \citealt{Jiang_2016, Wu_2022}), the Pan-Starrs 1 \citep{Morganson_2012,Banados_2023}, the Dark Energy 
Spectroscopic Instrument \citep{Yang_2023}, and the Hyper Suprime-Cam (HSC) Subaru Strategic Program \citep{Matsuoka_2016,Akiyama_2018, Matsuoka_2018,Niida_2020}.
Extensive efforts in spectroscopic follow-up observations have revealed that supermassive black holes (SMBHs)
have already formed and grown in mass to $M_\bullet \gtrsim 10^9~\msun$ within a short amount of cosmic time 
($\lesssim 1$ Gyr at $z\gtrsim 6$; \citealt{Wu_2015, Shen_2019,Wang_2021}, see also a recent review by \citealt{Fan_2023}).
This rapid growth of SMBHs represents one of the most intriguing unsolved mysteries in modern astrophysics.

The presence of such monster SMBHs has stimulated numerous ideas for their quick assembly mechanisms
\citep[e.g.,][]{Inayoshi_ARAA_2020,Volonteri_2021}: for instance, rapid gas feeding into the nuclei of early protogalaxies 
\citep{Volonteri_Rees_2005, Inayoshi_Haiman_Ostriker_2016, Toyouchi_2021}, the formation of massive seed BHs
through primordial star formation \citep{Omukai_2001, Bromm_Loeb_2003, Regan_2014, 
Inayoshi_2014, Wise_2019}, and runaway stellar collisions in dense clusters \citep{Devecchi_2009,Chon_Omukai_2020}.
Previous studies often focused on individual seeding scenarios within specific mass ranges and 
discussed their impact on subsequent evolution toward lower redshifts \citep{Bhowmick_2021,Spinoso_2023}.
However, in reality, the mass distribution of seed BHs is expected to be more continuous, ranging from so-called ``light seeds"
with $M_\bullet \sim 10^2~\msun$ to ``heavy seeds" exceeding $\sim 10^5~\msun$ due to the combination of various seeding mechanisms 
\citep[e.g.,][]{Sassano_2021,Toyouchi_2023}.
Recent studies by \citet{Li_2021} and \citet{Lupi_2021} have found that the formation of heavy seed BHs occurs more efficiently in 
rare, overdense regions of the universe, where nonlinear galaxy clustering creates peculiar external environments required for BH seeding
(e.g., H$_2$ dissociating radiation and dynamical heating via halo mergers).
Consequently, the black hole mass function (BHMF) for these seed populations is well approximated with a distribution of $\Phi_{M_\bullet} (\equiv dN/d\log M_\bullet )\propto M_\bullet^{-1}$ 
and extends the upper mass to $M_\bullet \gtrsim 10^5~\msun$ \citep{Li_2023a}.

The subsequent evolution of the BHMF is primarily influenced by the quasar luminosity function (QLF), which contains key information on 
the underlying physics on the radiative and accretion processes.
The shapes of both the BHMF and QLF have been extensively investigated by various approaches. 
Semi-analytical calculations have been employed to incorporate BH growth mechanisms, including gas accretion and BH mergers in the framework of 
hierarchical structure formation \citep{Shankar_2009,Ricarte_2018,Yung_2021,Kim_Im2021,Trinca_2022,Li_2023b}.
A recent work by \citet{Li_2023b} presented a theoretical model describing BH growth starting from initial seeding at $z \gtrsim 20$ to $z\sim 4$, 
taking into account the episodical accretion nature of quasars.
This model successfully explains the cosmic evolution of the observed $z \geq 4$ QLFs at UV absolute magnitude brighter than $-24$ mag
\citep[see][]{Akiyama_2018, Matsuoka_2018, Niida_2020} and furthermore offers predictions for the BHMF with masses ranging from 
the low-mass end at $M_\bullet \sim 10^{5-6}~\msun$ to $\sim 10^{10}~\msun$ at $4 \leq z\leq 11$.

Recently, the unprecedented sensitivity of the James Webb Space Telescope (JWST) has opened a new window to explore the low-luminosity 
active galactic nuclei (AGNs) at $z\sim 4-7$ \citep{Onoue_2023}.
Spectroscopic observations have confirmed the broad component of emission lines and the presence of AGNs 
\citep[e.g.,][]{Kocevski_2023, Harikane_2023_agn, Maiolino_2023_JADES,Larson_2023,Matthee_2023}.
The BH masses for those JWST-detected faint AGNs are $M_\bullet \sim 10^{6-8}~\msun$, which is 1-2 orders of magnitude lower than those of previously known quasars at $z>4$.
Intriguingly, some of the sources have extremely red colors in the rest-frame optical bands, while the rest-frame UV color seems fairly consistent with those of known 
Lyman-break galaxies at similar redshifts \citep[e.g.,][]{Kocevski_2023, Harikane_2023_agn,Barro_2023,Matthee_2023,Labbe_2023, Greene_2023}.
The so-called ``little red dots" sources are interpreted to be in a transition stage from a dust-obscured starburst to 
an unobscured luminous quasar by expelling gas and dust \citep{Fujimoto_2022}. 
These findings demonstrate the capability of JWST to extend high-$z$ BH studies to masses approaching the BH seed population,
while a significant 1-2 dex mass gap remains.

In this paper, we focus on UV-to-optical flares (in the rest frame) resulting from stellar tidal disruption events (TDEs) caused by BHs with masses around 
$\sim 10^{4-6}~\msun$ (\citealt{Rees_1988,Guillochon_2013,Stone_2013}; see also review papers by \citealt{vanVelzen_2020} and \citealt{Gezari_2021}).
Stellar TDEs can be triggered by low-mass BHs, where the BH gravitational force on the stellar surface dominates over the self-gravity
of plunging stars before they reach the event horizon.
These transient sources can be brighter than the persistent AGN emission, alleviating the restrictions imposed by the sensitivity of telescopes and 
thus improving the detectability of such low-mass BHs.
Moreover, it is worth noting that a large fraction of AGNs discovered with JWST exhibit extremely red colors and compact morphologies.
These characteristics suggest that the nuclear BHs are likely embedded in active star-forming regions, presumably dense stellar clusters, where TDEs occur frequently
due to short relaxation time.
The rate of TDEs, calculated based on the two-body relaxation time, is sensitive to the low-mass end of the BHMF around $M_\bullet \lesssim 10^6~\msun$
\citep{Magorrian_Tremaine_1999,Wang_Merritt_2004,Stone_Metzger_2016}.
Therefore, TDEs caused by low mass BHs can be used as a powerful tool to explore the underlying BH populations in the early universe 
\citep[e.g.,][]{Kashiyama_Inayoshi_2016}.

Motivated by the recent discovery of high-$z$ AGNs in the JWST surveys, we consider two types of transient TDEs triggered within unobscured and obscured AGNs.
TDEs occur more frequently in dense gaseous and stellar environments within dust-reddened obscured AGNs.
However, the emission from these TDEs is attenuated by dust grains, requiring deep JWST surveys for detection.
Since these surveys cover limited areas ($<1$ deg$^2$), TDEs produced by abundant BH populations with $M_\bullet \sim 10^{4-5}~\msun$ are detectable.
On the other hand, the wide-field surveys with Roman Space Telescope (RST; \citealt{Spergel_2015}) and Galaxy Reionization EXplorer and PLanetary Universe Spectrometer (GREX- PLUS; \citealt{GP_Inoue_2023})
are capable to identify rarer TDEs in unobscured AGNs.
Assuming the High-latitude Time Domain Survey with RST planned for supernova cosmology \citep{Rose_2021}, we expect to detect $\sim O(10)$ TDEs per year originating from AGNs 
with $M_\bullet \sim 10^{4-5}~\msun$ at $z\gtrsim 4$.

This paper is organized as follows. 
In Section~\ref{sec:LLAGN}, we begin by starting an overview of the current status of AGN observations with JWST.
We also provide a theoretical model to characterize the properties of galactic nuclei on scales of $\lesssim 100~\pc$ for JWST-detected AGNs at high redshifts. 
In Section~\ref{sec:TDE}, we give an estimate of the TDE rate led by BHs with $M_\bullet \simeq 10^{4-6}~\msun$ in those AGNs.
By convolving the TDE rate per galaxy with a theoretical BHMF model calibrated using the most up-to-date Subaru HSC AGN data \citep{Li_2023a, Li_2023b},
we make a prediction of the cosmic TDE rate as a function of BH mass and host galaxy properties including dust obscuration.
In Section~\ref{sec:obs_strategy}, we discuss the detectability of high-$z$ TDEs by the JWST, RST, and GREX-PLUS surveys, as well as lager sky-coverage 
surveys such as Euclid and the Legacy Survey of Space and Time (LSST) by the Vera C. Rubin Observatory, 
and propose survey strategies aimed at identifying transient high-redshift TDEs.
We finally summarize our findings and present some discussion in Section~\ref{sec:discussion}.

\begin{figure*}
\begin{center}
{\includegraphics[width=168mm]{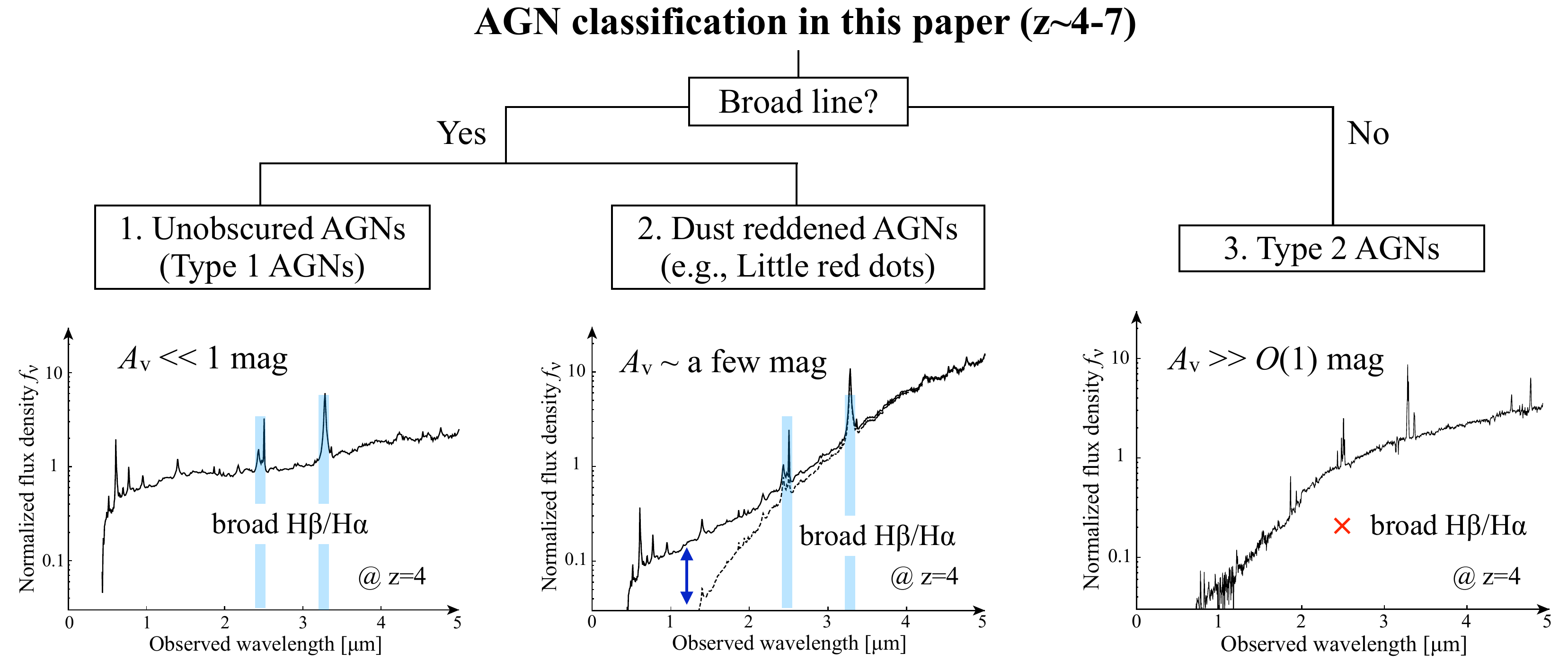}}
\caption{Flow chart of AGN classification in this paper;
(1) unobscured AGNs with broad components of the Balmer emission lines,
(2) broad-line AGNs but with dust extinction with $A_{\rm V}\sim $ a few,
and (3) AGNs without broad emission lines but heavily obscured by dust tori with $A_{\rm V}\gg O(1)$.
The left and right populations are classified as type 1 and 2 AGNs in the classical unified model picture \citep[e.g.,][]{Antonucci_1993,Urry_1995}.
The middle population (or ``little red dot") shows the unique spectral shape that can be explained by a dust-reddened AGN continuum (dashed)
coupled with excess emission in the rest-frame UV (indicated by the arrow).
The SED for each population is taken from (1) the composite quasar spectrum of \citet{VandenBerk_2001}, 
(2) the Torus model of \citet{Polletta_2006}, and (3) the type 2 Seyfert template of \citet{Francis_1991}, respectively.
The total AGN abundance at $z=4-11$ adopted in this paper is calibrated with those for the dust-reddened AGN population recently reported by JWST observations,
which is $1-2$ dex higher than the extrapolation of the UV luminosity function of more luminous quasars based on the ground-base surveys \citep[e.g.,][]{Niida_2020}. 
}
\label{fig:AGNclass}
\end{center}
\end{figure*}

\vspace{2mm}
\section{Low-luminosity AGNs at $z>4$\\ in the era of JWST}
\label{sec:LLAGN}

\subsection{Populations}
In this paper, we categorize AGNs into three groups based on their properties of the spectral energy distribution (SED) as summarized in Figure~\ref{fig:AGNclass}.
\begin{enumerate}
\item The first group comprises unobscured AGNs with broad emission lines (Type 1 AGNs).
High-$z$ unobscured AGNs observed with JWST exhibit the SED shapes consistent with a low-redshift composite spectrum of quasars \citep[e.g.,][]{VandenBerk_2001},
while their bolometric luminosities $L_{\rm bol}\sim 10^{43-45}{\rm erg~s}^{-1}$ are $\gtrsim 2$ dex fainter than those of the bright quasars 
previously known at similar redshifts \citep{Onoue_2023}. 
Broadband photometry and measurements of the narrow-line Balmer decrement suggest minimal to modest dust attenuation 
for this AGN population, with an average extinction value of $A_{\rm V}\simeq 0.3$ mag \citep[e.g.,][]{Maiolino_2023_JADES}.

\item An intriguing subset consists of the dust-reddened AGNs (or referred to as ``little red dots"), characterized by moderate obscuration
in the SED and compact morphology \citep{Barro_2023,Kocevski_2023,Harikane_2023_agn,Matthee_2023,Labbe_2023}.
This class of AGNs shows a distinctive spectral shape, which can be attributed to a dust-reddened AGN continuum
coupled with excess emission in the rest-frame UV produced either by unobscured stellar continuum of the host galaxy or scattered light from 
the buried AGN \citep[e.g.,][]{Zakamska_2005,Polletta_2006,Noboriguchi_2023}.
Despite significant extinction levels, typically with $A_{\rm V}\sim$ a few mag, these reddened sources show broad Balmer-line emission
indicating the presence of AGNs.

\item The third group includes AGNs lacking broad-line emissions due to substantial extinction (commonly referred to as Type 2 AGNs).
These AGNs are distinguished from ordinary galaxies by the presence of strong narrow-line emission features (e.g., [\ion{O}{3}])\footnote{
Some of narrow emission lines, e.g., [\ion{O}{3}], are also used as a characteristic of early protogalaxies with low metallicities 
\citep[e.g.,][]{Inoue_2011,Nakajima_Maiolino_2022}.} 
or high ionization emission lines 
(e.g., [\ion{Ne}{4}]$\lambda 2424$).
Meanwhile, the AGN continuum and broad-line emission at the nuclear scales are fully obscured by dust tori along our line of sight. 
To identify accreting BHs within such heavily obscured regions, alternative diagnostic methods are employed, including observations of  
X-rays \citep{Schramm_2013,Mezcua_2016,Baldassare_2015,Kawamuro_2019}, mid-infrared excesses \citep{Satyapal_2014,Delvecchio_2014}, radio core search \citep{Reines_2020,Ichikawa_2021}, and variability analyses \citep{Kokubo_2014,Baldassare_2018,Kimura_2020}.

\end{enumerate}

Deep NIRSpec observations have confirmed the presence of broad-line components in high-$z$ AGNs pre-selected or identified by JWST observations.
Based on measurements of the width and luminosity of the Balmer emission lines, as calibrated in the local universe, 
the virial masses of BHs powering these faint AGNs are estimated as $M_\bullet \sim 10^{6-7}~\msun$ for unobscured AGNs and $M_\bullet \sim 10^{7-8}~\msun$ 
for dust-reddened AGNs \citep[e.g.,][]{Kocevski_2023, Harikane_2023_agn,Matthee_2023,Maiolino_2023_JADES, Ubler_2023}.
These AGN populations, particularly those categorized into the first and second groups in Figure~\ref{fig:AGNclass}, place stringent constraints on the BHMF 
at early cosmic epochs and, consequently, the rate of TDEs.
However, Type 2 AGNs do not provide a robust measurement of BH mass, despite their potential prevalence in the overall AGN population. 
In the following discussion, we focus on two classes of AGNs: unobscured AGNs (Type 1) and obscured AGNs (dust-reddened AGNs).
Note that our analysis overlooks the contribution of TDE production in Type 2 AGNs, where the TDE emission might be heavily obscured as well. 
Regardless of the level of obscuration, the arguments presented below give a lower bound of the TDE rate and conservative detection forecasts.

\subsection{Abundances}

The first abundance measurement for high-$z$ faint AGNs was conducted by \cite{Onoue_2023}, using a single AGN candidate (CEERS-AGN-z5-1 or CEERS 2782), 
which has been confirmed as a broad-line AGN by \citet{Kocevski_2023}.
The AGN at $z\sim 5$ indicates an abundance as high as $\Phi_{\rm L} \sim 10^{-5}~\mpc^{-3}~{\rm mag}^{-1}$ at $\Muv \simeq -19.5$ mag within four pointings
covering a survey are of $34.5~{\rm arcmin}^2$.
This abundance is more than one order of magnitude higher than what was expected from the extrapolation of the QLFs derived from the ground-based surveys 
such as Subaru/HSC+SDSS \citep{Niida_2020} and CFHT Legacy Survey \citep{McGreer_2018}.
Subsequently, \citet{Harikane_2023_agn} and \citet{Maiolino_2023_JADES} reported 6 and 12 newly detected unobscured AGNs at $z \simeq 4-7$
from the CEERS (8 NIRSpec pointings covering 72~arcmin$^2$) and JADES survey fields (175~arcmin$^2$), respectively.
They found that $f_{\rm AGN}\sim 5-10\%$ of star-forming galaxies at $4<z<6$ (specifically, within JWST NIRSpec galaxy samples from each survey program)
exhibit signs of broad-line AGNs.
This high AGN fraction implies a prevalence of unobscured AGNs, corresponding to an abundance of 
$\Phi_{\rm L}\sim f_{\rm AGN} \Phi_{\rm L,\star}= 10^{-4}-10^{-3}~\mpc^{-3}~{\rm mag}^{-1}$,
where $\Phi_{\rm L,\star}$ denotes the UV galaxy luminosity function at the same redshift \citep[e.g.,][]{Finkelstein_2015,Bouwens_2021,Harikane_2022a}.
However, it is worth emphasizing that the selection function of NIRSpec targets can be quite complex and may vary across different observation tiers, 
particularly for slit spectroscopy modes influenced by pre-sample selection. 
As a result, constructing QLFs based on the AGN fraction within limited samples could introduce a level of unknown systematics.
In contrast, estimating the abundance of these faint AGNs based on the observed volume density yields a lower limit of
$\Phi_{\rm L} \sim 10^{-5}~\mpc^{-3}~{\rm mag}^{-1}$,
which aligns with other studies at similar redshifts and UV magnitudes.

Dust-reddened AGNs (or ``little red dots") have been identified in several survey fields \citep[e.g.,][]{Barro_2023,Labbe_2023}.
Among them, 31 sources have been confirmed as broad-line AGNs \citep{Kocevski_2023,Harikane_2023_agn,Matthee_2023,Greene_2023}.
The large number of detections highlights the significant abundance of these red and compact sources at $4<z<6$;
$\Phi_{\rm L} \simeq 10^{-5}~\mpc^{-3}~{\rm mag}^{-1}$ over $-18 <\Muv < -20$ \citep{Matthee_2023, Greene_2023}.
Moreover, BH mass measurements with spectroscopic observations of their broad emission lines put constraints on the BHMF at $z\simeq 4-5.5$ \citep{Matthee_2023}; 
$\Phi_{M_\bullet} \simeq 5\times 10^{-5}$ and $10^{-5}~\mpc^{-3}~{\rm dex}^{-1}$ at $M_\bullet \simeq 10^{7.5}$ and 
$10^{8.1}~\msun$, respectively.
Noteworthy, these abundances are approximately ten times higher than that for the type-1 AGN selected by SDSS and Subaru-HSC \citep{He_2023},
but seem consistent with the BHMF calculated with the continuity equation based on X-ray selected obscured samples that include the Type 2 AGNs with 
heavy extinction by dust and gas (\citealt{Ueda_2014}, see also \citealt{Small_Blandford_1992}).
This argument suggests that the obscured AGN samples with very red optical continuum and broad H$\alpha$ line could represent 
a bulk population of massive BHs at the end of cosmic reionization \citep{Li_2023b}.

\begin{figure}
\begin{center}
{\includegraphics[width=85mm]{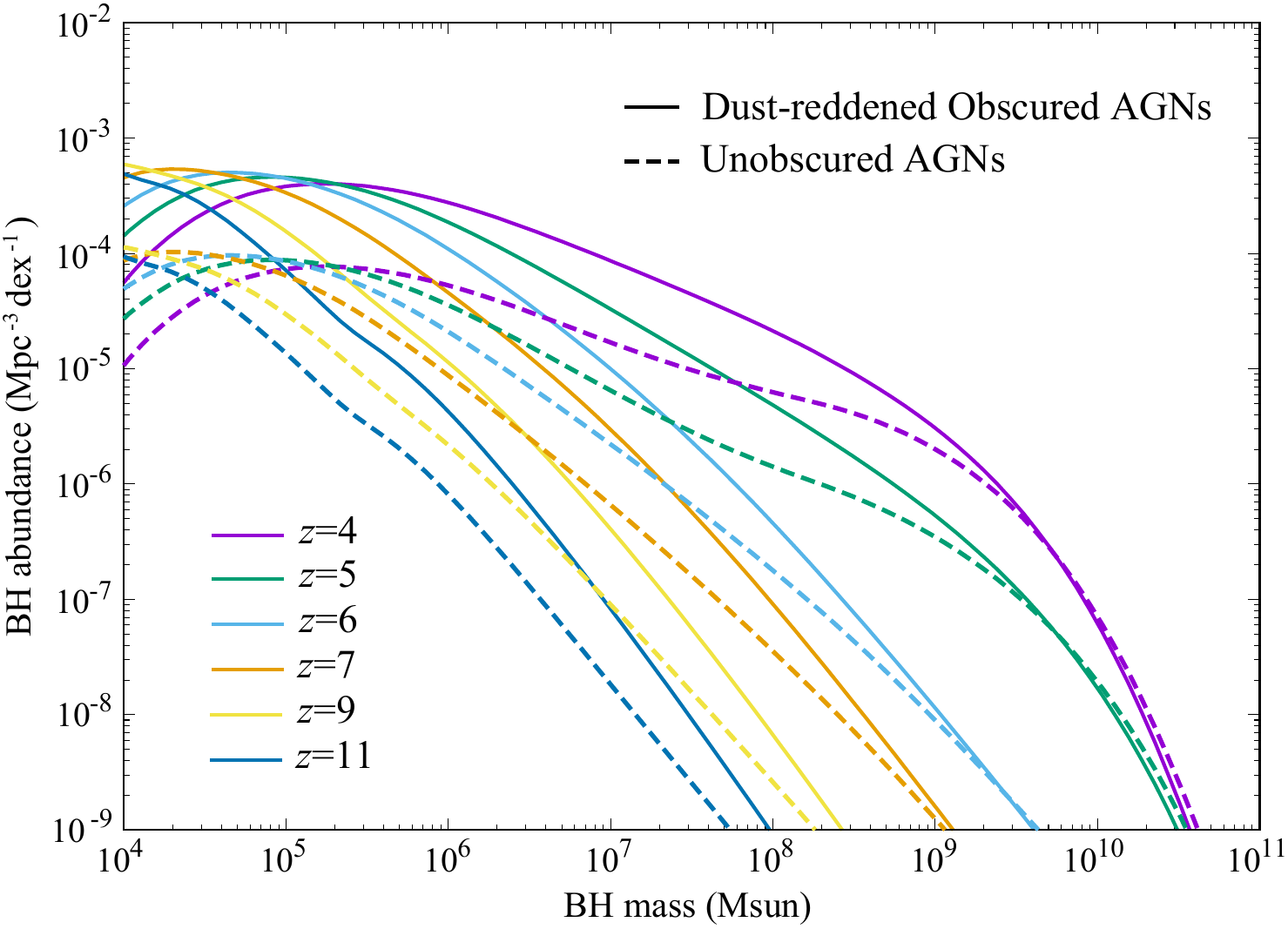}}
\caption{Black hole mass functions for the obscured (solid) and unobscured (dashed) population at $4\leq z \leq 11$,
constructed by a semi-analytical model in \cite{Li_2023a, Li_2023b}.
The cosmic evolution of the BHMF for high-$z$ quasars is modeled so that the observed unobscured QLFs at $4\lesssim z \lesssim 6$ is well described.}
\label{fig:MF}
\end{center}
\end{figure}

The recent theoretical models of the QLF and BHMF evolution have also suggested a high AGN abundance at fainter ends \citep[e.g.,][]{Trinca_2022,Li_2023a, Li_2023b}.
In Figure~\ref{fig:MF}, we show the BHMFs for the obscured (solid) and unobscured (dashed) population at $4\leq z \leq 11$
constructed by \cite{Li_2023a, Li_2023b}\footnote{We adopt the model where all the seed BHs formed at $z>15$, with a number density of $n_{\rm seed} \simeq 10^{-3}~{\rm Mpc}^{-3}$, 
have subsequent growth to participate in the assembly of SMBHs and end up in quasar host galaxies by $z\simeq 6$. 
\citet{Li_2023b} demonstrated that this model successfully reproduce the high abundance of JWST-detected AGNs at fainter ends.}, 
where the BH growth model is calibrated to reproduce the redshift-dependent QLFs for unobscured populations \citep[e.g.,][]{Akiyama_2018, Matsuoka_2018,Niida_2020,Matsuoka_2023}.
The BHMF evolves to the high-mass end between the BH seeding epochs of $z>15$ and the end of cosmic reionization at $z\sim 5$.
Toward a lower redshift of $z\sim 4$, the growth of substantially heavy BHs with $M_\bullet > 10^9~\msun$ is stunted and makes the shape of the distribution
consistent with the observational results.
The abundance at the low-mass end of $10^5\lesssim M_\bullet/\msun \lesssim 10^7$, where massive BHs trigger stellar tidal disruption events
(see Section \ref{sec:TDE}), monotonically increases and reaches $\Phi_{M_\bullet} \simeq 10^{-4}$ and $10^{-3}~\mpc^{-3}~{\rm dex}^{-1}$
for the unobscured and obscured BH population, respectively.
The number difference between the two populations reflects the nature of obscuration due to dust and gas in the galactic nuclei.
In this work, we adopt the obscuration fraction fitted by \citet{Ueda_2014} based on X-ray AGN observations up to $z\sim 5$.

It is worth emphasizing that the BHMF model we adopt in this paper considers BH populations formed in relatively biased regions
of the universe with mass variance of $\gtrsim 3\sigma$, where the quasar progenitor halos are likely irradiated by intense
H$_2$-photodissociating radiation from nearby star-forming galaxies and heat the interior gas by successive mergers 
\citep[e.g.,][]{Wise_2019,Li_2021, Lupi_2021} and thus the mass distribution of seed BHs is extended to $M_\bullet \sim 10^5~\msun$ \citep{Toyouchi_2023, Li_2023a}. 
However, their model neglects contributions from substantially low-mass BH populations and their mass growth in less massive halos
formed in the typical regions of the early universe, leading to underestimate of the BH abundance at the low-mass end of $M_\bullet < 10^7~\msun$.
Despite this caveat, our results discussed in the following sections manifest the importance of transient observations with JWST and RST to probe the low-mass BH population
and thus to constrain their seeding mechanisms \citep[e.g.,][]{Inayoshi_ARAA_2020, Volonteri_2021}.

\subsection{Dusty Nuclear Star Clusters}
\label{sec:cluster}

The unprecedented infrared sensitivity and spatial resolution of JWST have found a remarkable fact that JWST-detected obscured AGNs 
produce a sufficient amount of dust grain within a compact region with size of $R_{\rm e}\sim 100~\pc$ \citep{Matthee_2023, Labbe_2023}.
In particular, those red sources found in the UNCOVER field are well resolved down to $R_{\rm e}\sim 10~\pc$
because of a high magnification in the gravitational lensing field and the median value of the half-light radius of the sample is 
$\langle R_{\rm e}\rangle \sim 50~\pc$ \citep{Labbe_2023}.
From the SED fitting analysis, the visual extinction to explain the continuum slope is considered to be 
as high as $A_{\rm V}\sim 2-4~{\rm mag}$ \citep{Harikane_2023_agn, Labbe_2023, Greene_2023}; for instance, a sample of the red sources spectroscopically 
confirmed as an AGN indicates a strong dust attenuation with an average $\langle A_{\rm V}\rangle \simeq 3$ mag and a range spanning $1.2-4.6$ 
\citep{Matthee_2023, Furtak_2023b}.
Adopting a dust-to-gas mass ratio of $\mathcal{D}= 0.01$, the hydrogen column density along the line of sight is calculated as
$N_{\rm H}\sim 5.6\times 10^{21}~{\rm cm}^{-2}(A_{\rm V}/3.0)(D/0.01)^{-1}$.
Thus, the dust mass spherically distributed within $R_{\rm e}$ is given by
\begin{align}
M_{\rm dust} & \simeq \frac{4\pi}{3} g(\gamma) \mathcal{D} \mu m_{\rm H} N_{\rm H} R_{\rm e}^2, \nonumber\\
& \simeq 6.1\times 10^3~\msun \left(\frac{A_{\rm V}}{3.0}\right) \left(\frac{R_{\rm e}}{50~\pc}\right)^2,
\end{align}
where $m_{\rm H}$ is the hydrogen mass, $\mu$ is the mean molecular weight, and the gas density is assumed to follow 
$\propto r^{-\gamma}$ with $\gamma<1$ and thus $g(\gamma) = (1-\gamma)/(1-\gamma/3) \sim 1$.

Dust production in galaxies is led by supernova explosions of short-lived massive stars and stellar wind from long-lived asymptotic 
giant branch stars \citep[e.g.,][]{Asano_2013}. 
In galaxies at $z\gtrsim 6$ with ages of $\lesssim 1~{\rm Gyr}$, the dust-to-stellar mass ratio is expected to be 
\begin{equation}
f_{\rm dust,\star} = \frac{M_{\rm dust}}{M_\star} \sim
\begin{cases}
1.5 \times 10^{-5},\\
5.0 \times 10^{-4},
\end{cases}
\end{equation}
where the upper and lower value correspond to the case with continuous and instantaneous star formation, respectively \citep{Valiante_2009}.
We note that the exact value, in general, depends on the metallicity and shape of the stellar initial mass function, 
but the scatter becomes smaller at $t\gtrsim 1$ Gyr.
In the following, we adopt the ratio of $f_{\rm dust,\star} \sim 1.5\times 10^{-5}$ for the continuous star formation history and 
estimate the stellar mass in the nuclear region as 
\begin{equation}
M_{\rm \star, nuc} \simeq 4.1\times 10^8~\msun 
\left(\frac{A_{\rm V}}{3.0}\right) 
\left(\frac{R_{\rm e}}{50~\pc}\right)^2.
\label{eq:M_star_nuc}
\end{equation}
This estimate also leads to a stellar surface density of $\Sigma_{\star, \rm nuc}\simeq 5.2\times 10^4~\msun~\pc^{-2}$,
which is comparable to or lower lower than an upper stellar density limit for the densest star clusters or the densest elliptical galaxy progenitors 
(\citealt{Hopkins_2010}; see also \citealt{Vanzella_2023,Baggen_2023}).
Assuming the stellar mass density to follow a power-law profile of $r^{-2}$, it is expressed by 
\begin{align}
\rho_\star(r) 
\simeq 6.5\times 10^5 ~\zeta~\msun~\pc^{-3}
\left(\frac{r}{\pc}\right)^{-2},
\end{align}
where the value of $\zeta$ is defined by
\begin{equation}
\zeta \equiv \left(\frac{A_{\rm V}}{3.0}\right) \left(\frac{R_{\rm e}}{50~\pc} \right)
\end{equation}
as this combination appear in the following equations.
Note that in this assumption, the stellar mass distribution is concentrated toward the center compared to the distribution of gas and dust (see Eq.~\ref{eq:M_star_nuc}).
If we approximate the stellar density profile as a singular isothermal sphere, $\rho_\star = \tilde{\sigma}^2/(2\pi Gr^2)$,
the stellar velocity dispersion of the cluster is given by 
\begin{align}
\tilde{\sigma} = \sqrt{\frac{GM_{\star, \rm nuc}}{2R_{\rm e}}}
\simeq 132~\zeta^{1/2}~\kms 
\label{eq:sigma-nuc}
\end{align}
at $r\gtrsim r_{\rm h}\equiv  GM_{\bullet}/\tilde{\sigma}^2$.
For our fiducial case, the BH gravitational influence radius is give by
\begin{align}
r_{\rm h} \simeq 0.25~\zeta^{-1}~\pc \left(\frac{M_{\bullet}}{10^6~\msun}\right)
\end{align}
We note that the BH influence radius is smaller than the half-light radius (i.e., $r_{\rm h}<R_{\rm e}$)
when the BH mass is as low as $M_\bullet <2.0\times 10^8~\msun (A_{\rm V}/3.0) (R_{\rm e}/50~\pc)^2$.

It is also worth pointing out that Eqs.~(\ref{eq:M_star_nuc}) and (\ref{eq:sigma-nuc}) implies that a correlation between the stellar mass and velocity dispersion as
\begin{align}
M_{\star, \rm nuc} = 1.3\times 10^8~\msun 
\left(\frac{A_{\rm V}}{3.0}\right)^{-1} 
\left(\frac{\tilde{\sigma}}{100~\kms}\right)^{4}.
\end{align}
The scaling relation of $M_{\star, \rm nuc}\propto \tilde{\sigma}^4$ is consistent with that seen for nearby massive galaxies \citep[e.g.,][]{Kormendy_Ho_2013},
and the normalization of the relation agree with the mass of nuclear star clusters in the local universe 
\citep[see the observed data summarized in Figure~1 of][]{Stone_2017b}.

\begin{figure}
\begin{center}
{\includegraphics[width=84mm]{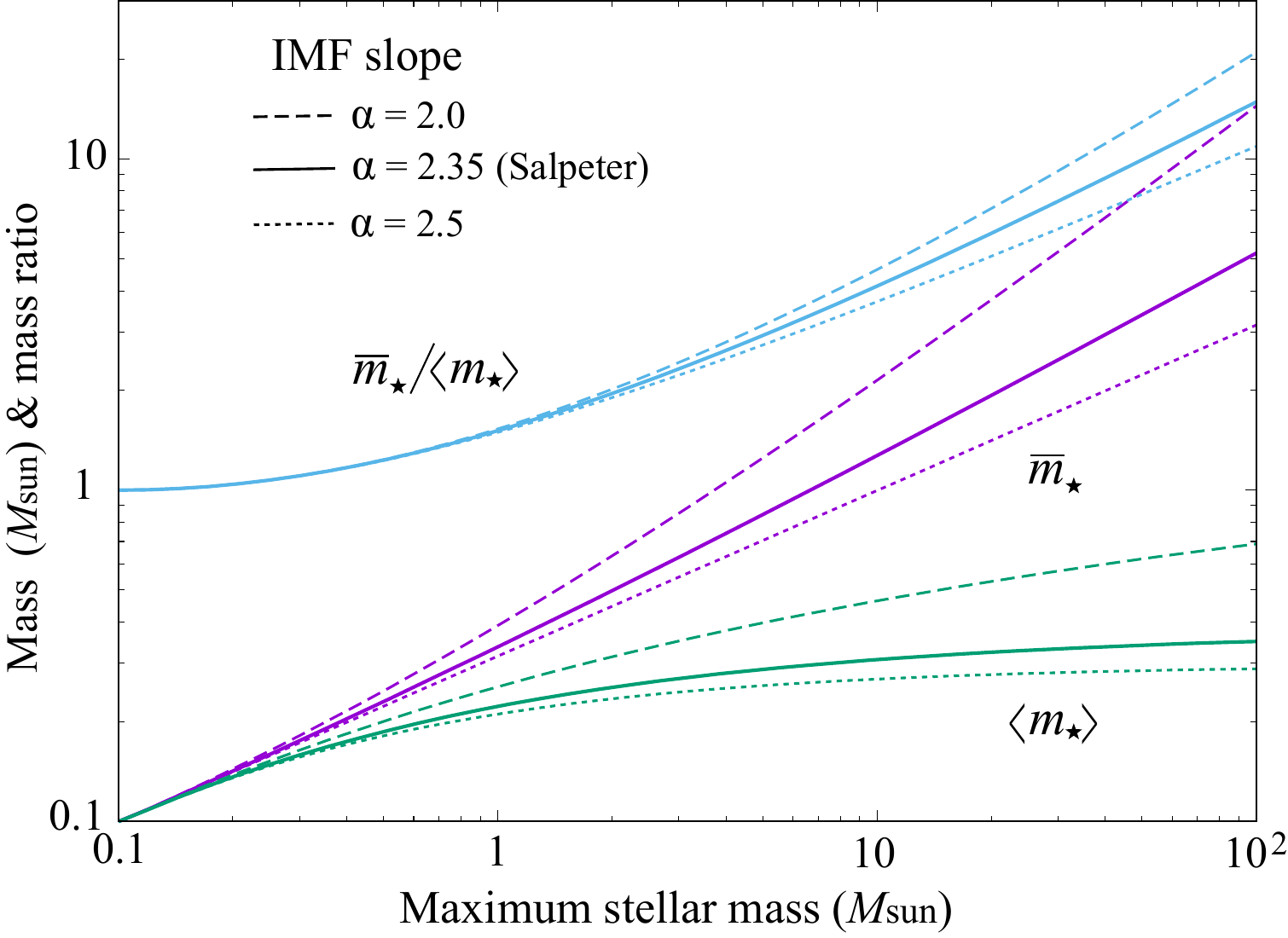}}
\caption{The mean mass $\langle m_\star \rangle$, the effective mass $\bar{m}_\star$, and their ratio
as a function of the maximum mass of a stellar mass distribution with a power-law slope of $\alpha$.
}
\label{fig:simf}
\end{center}
\end{figure}

\vspace{2mm}
\section{Stellar Tidal Disruption in high-redshift AGNs}\label{sec:TDE}

\subsection{Timescales}\label{sec:time}

Let us consider the fate of a population of ordinary low-mass stars, massive main-sequence stars, and stellar remnant BHs in the dense nuclear star cluster
surrounding a massive BH in a red obscured AGN.
Here, we assume a spherical shape of the cluster and estimate the two-body relaxation time for the low-mass stars as
\begin{align}
t_{\rm rel} & = 0.34~\frac{\sigma_\star^3}{G^2\rho_{\rm \star} \bar{m}_\star  \ln \Lambda},\nonumber\\[5pt]
& \simeq 99.4~\zeta^{-3/2}~{\rm Myr}
\left(\frac{M_\bullet }{10^6~\msun}\right)^{2}
\left(\frac{\bar{m}_\star}{\msun}\right)^{-1}
\left(\frac{r}{r_{\rm h}}\right)^{1/2},
\label{eq:trelax}
\end{align}
where the stellar velocity dispersion in the nucleus at $r<r_{\rm h}$ is 
$\sigma_\star\simeq 147~\kms (M_\bullet/10^6~\msun)^{1/2}(r/0.1~\pc)^{-1/2}$,
$\ln \Lambda = \ln (M_\bullet/\langle m_\star \rangle)\sim 14$, $\langle m_\star \rangle$ is the mean stellar mass,
and the effective mass, $\bar{m}_\star \equiv \langle m_\star^2 \rangle/\langle m_\star \rangle$, is the ratio of the mean-squared stellar mass to the mean stellar mass.
Figure~\ref{fig:simf} presents the mean mass, the effective mass, and their ratio as a function of the maximum mass of a stellar mass distribution 
with a power-law slope of $\alpha$; $dn_\star \propto m_\star^{-\alpha} dm_\star$ with $m_{\rm min}=0.1~\msun$.
While the mean mass weakly depends on $m_{\rm max}$ for $\alpha \geq 2$, the effective mass increases with $m_{\rm max}$ following $m_{\rm max}^{3-\alpha}$ 
in the limit of $m_{\rm max} \gg m_{\rm min}$ when $2<\alpha <3$.
For a Salpeter mass distribution with $\alpha=2.35$ and $m_{\rm min}=0.1~\msun$, one obtains $\bar{m}_\star \simeq 1.3~(5.2)~\msun$ for $m_{\rm max}=10~(100)~\msun$.
Therefore, the presence of massive perturbers in the cluster region accelerates the relaxation process 
compared to the case where all stars have a single mass.

\begin{figure}
\begin{center}
{\includegraphics[width=85mm]{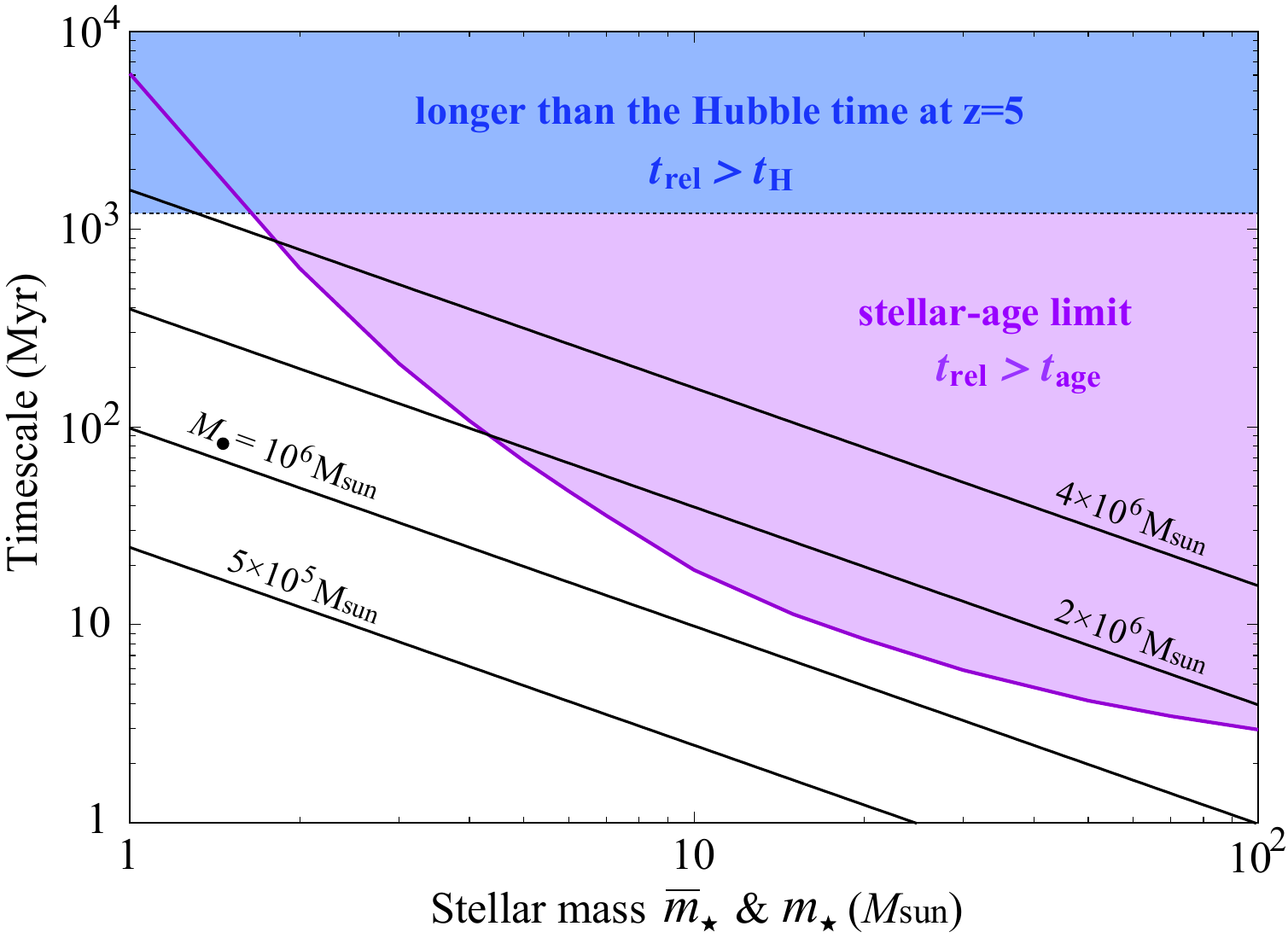}}
\caption{Dynamical friction timescales as a function of the effective mass $\bar{m}_\star$ of stars
surrounding a massive BH with $M_\bullet = 5\times 10^5$, $10^6$, $2\times 10^6$, and $4\times 10^6~\msun$ (from the bottom to the top; black lines).
The dynamical friction timescale is evaluated at the BH influence radius of $r_{\rm h}$ for $\zeta=1$ and scales with $\zeta^{-3/2}$ (see Eq.~\ref{eq:trelax}).
The stellar age of a star with $m_\star$ (magenta curve) and the Hubble time at $z=5$ are shown, respectively. 
}
\label{fig:time}
\end{center}
\end{figure}

Figure~\ref{fig:time} shows the dynamical friction timescale as a function of the effective mass $\bar{m}_\star$ of stars surrounding a massive BH with 
$M_\bullet= 5\times 10^5$, $10^6$, $2\times 10^6$, and $4\times 10^6~\msun$ (from the bottom to the top; black lines).
The timescale is evaluated at the BH gravitational influence radius of $r=r_{\rm h}$.
We also overlay the age of a main sequence star as a function of mass for $Z=0$ taken from \citet{Marigo_2001}
(note that the stellar age becomes longer as the stellar metallicity increases).
By equating $t_{\rm age}(m_{\rm max,TDE})=t_{\rm rel}[\bar{m}_\star(m_{\rm max,TDE})]$, one can calculate the upper mass of stars that can feed the nuclear BH within the lifetime.
In a stellar cluster hosting a BH with $M_\bullet \simeq 10^6~\msun$, the upper mass is estimated as $m_{\rm max,TDE}\simeq 3.5~\msun$, 
corresponding to $\bar{m}_\star =0.7~\msun$ and $t_{\rm rel}=150$ Myr.
Therefore, in the cluster, B-type stars with $m_\star \lesssim 3.5~\msun$ can sink to the center and lead to TDEs within the main-sequence stage.
The upper mass rapidly increases as the BH mass decreases;
for instance, $m_{\rm max,TDE}\simeq 10~\msun$ for $M_\bullet = 5\times 10^5~\msun$ and 
$m_{\rm max,TDE}\gg 100~\msun$ for $M_\bullet \lesssim 3\times 10^5~\msun$ as the stellar age hardly depends on the stellar mass at $m_\star >50~\msun$.
In conclusion, for a dense cluster in galactic nuclei hosting a BH with $M_\bullet \lesssim 5\times 10^5~\msun$, 
massive OB stars can contribute to the occurrence of TDEs without undergoing supernova explosions or gravitational collapse to BHs,
unlike TDEs observed in the low-redshift universe.
The role of low-mass BHs at the high-$z$ universe as factories of massive stellar TDEs has been proposed in \citet{Kashiyama_Inayoshi_2016},
in a context of the direct-collapse BH formation during rapid assembly of infalling gas and stars within protogalaxies in overdense regions of the universe.

\subsection{Tidal disruption event rates}

An SMBH tidally disrupt a main-sequence star of mass with $m_\star$, if the pericenter distance of the stellar orbit is
close or located inside the tidal radius $r_t$, defined as
\begin{align}
r_t &= \left(\eta^2 \frac{M_\bullet}{m_\star}\right)^{1/3} r_\star,\\
&\simeq 0.49~{\rm AU}~\eta^{2/3}
\left(\frac{M_\bullet}{10^6~\msun}\right)^{1/3}
\left(\frac{m_\star}{\msun}\right)^{-1/3}
\left(\frac{r_\star}{\rsun}\right), \nonumber
\end{align}
where $r_\star$ is the stellar radius of interest.
The quantity of $\eta$ characterizes the ratio between the duration of periapsis passage at the tidal radius and 
the hydrodynamical timescale of the star, and thus depends on the internal stellar structure.
For Sun-like stars and massive stars, the equation of state at the interior is approximated with a polytropic law
and thus we set $\eta \simeq 0.84$.
A star at a distance of $r$ from the cluster center will be removed if the orbital angular momentum is small enough
owing to tidal disruption.
The angular size $\theta_{\rm lc}$ of a stellar velocity vector that populates itself into 'loss cone' is given by
\begin{align}
\theta_{\rm lc} &\simeq \left(\frac{r_{\rm t}}{r}\right)^{1/2}\nonumber\\
&\simeq 2.9\times 10^{-3}~\zeta^{1/2} \left(\frac{M_\bullet}{10^6~\msun}\right)^{-1/3}
\left(\frac{r}{r_{\rm h}}\right)^{-1/2}, 
\end{align}
at $r\lesssim r_{\rm h}$ and the angular size decreases outward as $\theta_{\rm lc} \propto r^{-1}$ at $r>r_{\rm h}$, 
where the stellar gravity dominates over the BH gravity.

\begin{figure*}
\begin{center}
{\includegraphics[width=130mm]{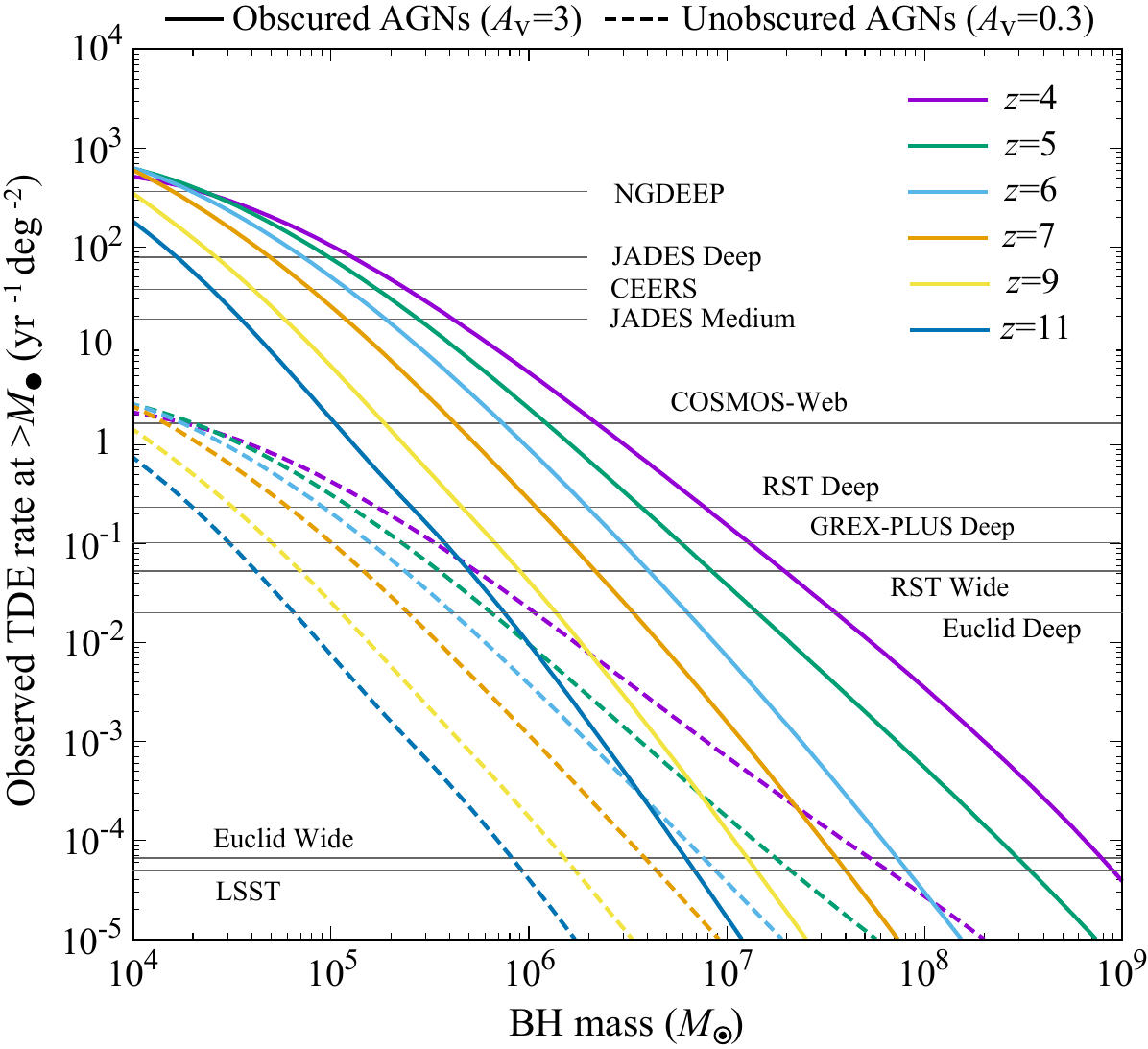}}
\caption{The predicted rate density (in units of yr$^{-1}$~deg$^{-2}$) of stellar TDEs within obscured (solid) and unobscured (dashed) AGNs within the redshift range of $4\leq z \leq 11$.
The horizontal lines represent the inverse of the survey area in one-year period planned for individual JWST missions 
(COSMOS-Web, CEERS, JADES, and NGDEEP), the High-latitude Time Domain Survey for supernova cosmology with RST, GREX-PLUS, Euclid, and LSST. 
The TDE rate in the obscured AGNs ($A_{\rm V}\simeq 3$) is approximately two orders of magnitude higher than that for unobscured BH AGNs ($A_{\rm V}\simeq 0.3$), where we fix $R_{\rm e}=50~\pc$ and $\bar{m}_\star/\langle m_\star \rangle =4$ (for a Salpeter stellar mass function with $0.1\leq m_\star/\msun \leq 10$).
These findings highlight the capability of deep JWST observations to detect TDEs in obscured AGNs with low-mass BHs, while the wider coverage of the RST, Euclid, GREX-PLUS, and LSST surveys 
is pivotal for identifying TDEs in unobscured AGNs.}
\label{fig:tde_rate}
\end{center}
\end{figure*}

Using the simplified formulation of \citet{Syer_Ulmer_1999}, we estimate the stellar TDE rate as 
$\dot{N}_{\rm TDE}=\dot{N}_<+\dot{N}_>$, where the two terms are the stellar consumption rates in the empty and full loss cone regimes, respectively.
At the scale of interest at $r\sim r_{\rm h}$, the total rate is dominated by the first term and the 
rate per star is given by
\begin{align}
\frac{d\dot{N}_<}{dN_\star} = \frac{1}{t_{\rm rel} \ln (2/\theta_{\rm lc})},
\end{align}
and the integrated rate is expressed by
\begin{align}
\dot{N}_< &  
=4\pi \int_0^{r_0} \frac{\ln \Lambda_{\rm lc}G^2 \rho_\star^2(r)r^3}{k \sigma_\star^3(r)} ~\frac{\bar{m}_\star}{\langle m_\star \rangle}~\frac{dr}{r}\nonumber\\[5pt]
& = \frac{4\sqrt{2} \ln \Lambda_{\rm lc}}{k \pi}\frac{\tilde{\sigma}^3}{GM_\bullet}~\frac{\bar{m}_\star}{\langle m_\star \rangle} \left(\frac{r_0}{r_{\rm h}}\right)^{1/2},
\end{align}
where $k=0.34$ is a numerical prefactor from the relaxation time, and $\ln \Lambda_{\rm lc} \equiv \ln \Lambda/\ln (2/\theta_{\rm lc})$.
The critical radius $r_0$ is the location where the per-star consumption rate is equally partitioned between 
the full and empty loss cone regimes, and is approximately given by solving the equation as
\begin{align}
\left(\frac{r_0}{r_{\rm h}}\right)^3= \frac{k\pi}{2\ln \Lambda_{\rm lc}(r_0)}\cdot \frac{r_{\rm t}M_\bullet}{r_{\rm h}\langle m_\star \rangle}
\Rightarrow r_0 \simeq 1.3 \zeta^{1/3}r_{\rm h}
\end{align}
Since the scaling relations we have described remain valid at $r\lesssim r_{\rm h}$, we evaluate $\dot{N}_<$ at $r=r_{\rm h}$
and neglect the contribution from the full loss-cone regime at $r>r_0$.
Therefore, we give an estimate of the TDE rate per galaxy as
\begin{equation}
\dot{N}_{\rm TDE}  \simeq 2.8\times 10^{-2}~\zeta^{5/3}~{\rm yr}^{-1} 
\left(\frac{\bar{m}_\star}{4.0~\langle m_\star \rangle}\right)\left(\frac{M_\bullet }{10^6~\msun}\right)^{-1}.
\label{eq:NTDE}
\end{equation}
where we assume $\bar{m}_\star/\langle m_\star \rangle = 4.0$, corresponding to the case for a Salpeter mass distribution with a mass range of $0.1-10~\msun$ (see Figure~\ref{fig:simf}).
Note that this ratio can be higher if the mass of the nuclear BH is below $\simeq 5\times 10^5~\msun$ (see discussion in Section~\ref{sec:time}).
While the presence of massive stellar perturbers in the cluster region enhances the TDE rate, the fraction of TDEs caused by those massive stars would be reduced from the estimate in Eq.~(\ref{eq:NTDE})
by a factor of $t_{\rm age}/t_{\rm AGN}\sim 0.1$, where $t_{\rm age}\sim 10$ Myr represents the typical age of massive stars and $t_{\rm AGN}\simeq 100$ Myr is the typical AGN lifetime \citep[e.g.,][]{Martini_2004}.

The TDE rate depends on the properties of the obscured nuclear region as $\dot{N}_{\rm TDE} \propto (A_{\rm V} R_{\rm e})^{5/3}$.
The rate estimated for high-redshift dust obscured AGNs is indeed consistent with the rate expected in low-redshift ultra-luminous infrared galaxies \citep{Tadhunter_2017}.
Furthermore, the expected TDE rate for less obscured systems is reduced to $\gtrsim 10^{-5}-10^{-4}~{\rm yr}^{-1}~{\rm galaxy}^{-1}$ for $A_{\rm V}\simeq 0.1-0.4$ mag,
which generally agrees to the rate for the field galaxy and post-starburst galaxy population (e.g., \citealt{vanVelzen_2014, French_2016}; see also \citealt{Gezari_2021}).

It is also worth noting that two-body relaxation in spherical star clusters is generally considered to give a conservative floor for the true TDE rate.
Other effects may contribute to the relaxation rate in galactic nuclei and potentially enhance the TDE rate; for instance, the non-spherical potential of the stellar cluster,
and resonant relaxation \citep[e.g.,][references therein]{Alexander_2012,Merritt_2013_book}.
Of particular significance, the interaction between AGN disks and stars, involving mechanisms such as gas dynamical friction, orbital migration, and the excitation of 
multi-body stellar scattering, plays an essential role in triggering TDEs \citep{Ryu_2023,Prasad_2023,Wang_2023}.
Additionally, strong gravitational interactions induced by a massive binary BH can accelerate the process and enhance the TDE rate
compared to the standard relaxation process around a single BH \citep[e.g.,][]{Chen_2011}.

\vspace{2mm}
\subsection{Cosmic TDE rates}
To estimate the TDE rate density in a given survey area, we convolve the rate per galaxy with the BHMFs at various redshifts (see Figure~\ref{fig:MF}) as 
\begin{equation}
\mathcal{R}_{\rm TDE} (>M_\bullet)= \int_{M_\bullet}^{\infty} 
\dot{N}_{\rm TDE} ~\frac{\Phi_{M_\bullet}(M'_\bullet)}{\ln 10}~\frac{\Delta V_{\rm c,deg}}{1+z} ~\frac{dM'_{\bullet}}{M'_{\bullet}},
\end{equation}
where $\Delta V_{\rm c,deg}$ is the comoving volume between $z-0.5$ and $z+0.5$ within a deg$^2$ solid angle and the $(1+z)$ factor 
is required to convert the rate in the galaxy-rest frame to the observer frame.
In Figure~\ref{fig:tde_rate}, we show the TDE rate density (in units of yr$^{-1}$ deg$^{-2}$) as a function of BH mass 
for the obscured BH population ($A_{\rm V}=3$ mag; left) and unobscured one ($A_{\rm V}=0.3$ mag; right).
Overall, the TDE rate is higher at lower BH masses because (1) the event rate per galaxy is higher following $\dot{N}_{\rm TDE}\propto M_\bullet^{-1}$ and 
(2) lower-mass SMBHs are more abundant (see Figure~\ref{fig:MF}).
At higher redshifts ($z\gtrsim 7$), the TDE rate increases toward the lower mass following $\mathcal{R}_{\rm TDE} \propto M_\bullet ^{-1.5}$ 
down to $M_\bullet \sim 10^5~\msun$.
As the redshift is lowered to $z\sim 4$, the distribution is extended to the higher BH mass regime following the cosmic evolution of the entire BH population.
Furthermore, the TDE rate for the obscured BH population is approximately two orders of magnitude higher than that for the unobscured BH population,
reflecting that dust-reddened, obscured AGNs are more abundant and lead to TDEs more frequently.

The horizontal lines represent the inverse of the survey area within a one-year timeline for individual programs with JWST, RST, GREX-PLUS, Euclid, and LSST.
In this paper, we consider the following four deep imaging surveys using JWST NIRCam: the JWST Advanced Deep Extragalactic Survey 
(JADES; \citealt{Eisenstein_2023}), the Cosmic Evolution Early Release Science (CEERS) survey \citep{Finkelstein_2023}, 
the COSMOS-Web survey \citep{Casey_2023}, and the Next Generation Deep Extragalactic Exploratory Public (NGDEEP) survey \citep{Bagley_2023}. 
For the RST survey, we adopt the initial survey design for the High-latitude Time Domain Survey for supernova cosmology \citep[e.g.,][]{Rose_2021},
which employs the wide and deep tiers with a survey area of $19.04$ and $4.20$ deg$^2$, respectively.
For the GREX-PLUS survey, a deep imaging survey survey with an area of $10$ deg$^2$ is considered \citep{GP_Inoue_2023}.
Additionally, Euclid, launched successfully  in July 2023, is planned to conduct two major surveys: a wide survey of $\sim 15,000~{\rm deg}^2$
and a deep survey of $\sim 50~{\rm deg}^2$, covering substantially larger sky areas compared to the survey programs mentioned above
\citep{Laureijs_2011,Euclid_I_2022}\footnote{https://www.cosmos.esa.int/web/euclid}.
The LSST ten-year survey will involve more than five million exposures, collecting data to produce a deep, time-dependent, panorama of 
$\sim 20,000~{\rm deg}^2$ of the sky \citep{Ivezic_2019}\footnote{https://www.lsst.org/scientists/survey-design}.

In the case of obscured AGN populations, the TDE rate is sufficiently high and thus a substantial number of events are expected within the surveyed area. 
However, it is crucial to note that when considering the detection limit, observations of TDEs within obscured AGN populations should be restricted 
to a relatively small field of view achievable through deep JWST observation programs, where the limiting magnitude should be at or 
deeper than 28 mag to ensure reliable detection.
Conversely, for unobscured AGN populations, the occurrence rate of TDEs is low. 
Therefore, wide-field ($\gtrsim 2~{\rm deg}^2$) surveys with imaging with RST and GREX-PLUS deeper than $25-26$ mag are required to effectively identify
identifying TDEs in these populations (see more details in Section~\ref{sec:obs_strategy}).

\section{Observational Signatures}
\label{sec:obs_strategy}

The accretion luminosity of TDEs occurred in high-$z$ AGNs can be sufficiently high and thus the associated emission would be detectable 
even from $z\gtrsim 4-7$.
The fallback accretion rate is estimated as 
\begin{equation}
\dot{M}_{\rm fb}\simeq \frac{m_\star}{3t_{\rm fb}}\left(\frac{t}{t_{\rm fb}}\right)^{-5/3},
\end{equation}
where the fallback time $t_{\rm fb}$ is calculated as
\begin{align}
t_{\rm fb} &\simeq 0.11~{\rm yr}
\left(\frac{M_\bullet}{10^6~\msun}\right)^{1/2}
\left(\frac{m_\star}{\msun}\right)^{-1}
\left(\frac{R_\star}{\rsun}\right)^{3/2},
\end{align}
\citep{Stone_2013} and its analytical expression $t_{\rm fb}\propto M_\bullet^{1/2}$ fits well the relation between the light curve 
decay timescale and the BH mass for optical/UV-selected TDEs \citep[e.g.,][]{vanVelzen_2020}.

\begin{figure*}
\begin{center}
{\includegraphics[width=88mm]{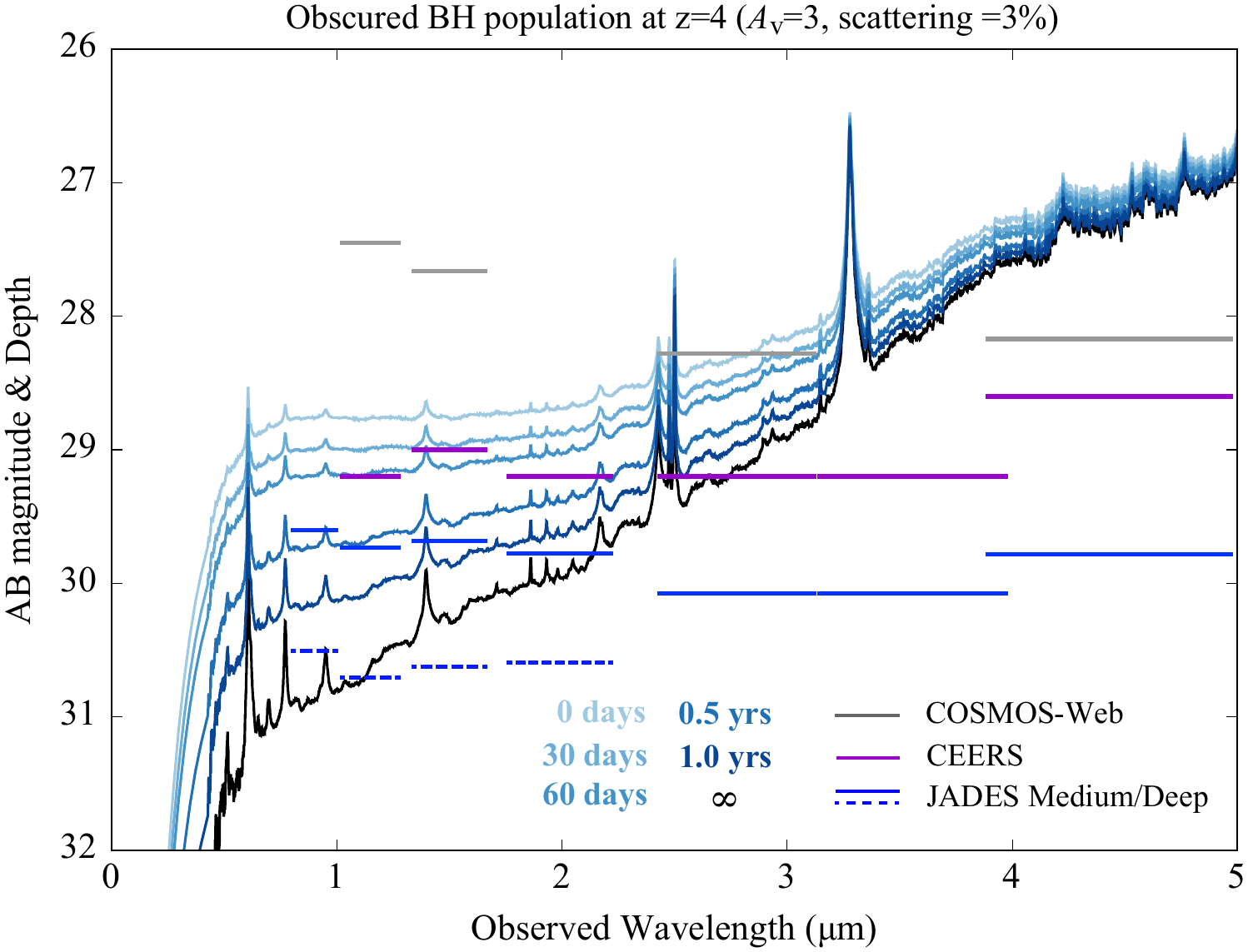}}\hspace{3mm}
{\includegraphics[width=88mm]{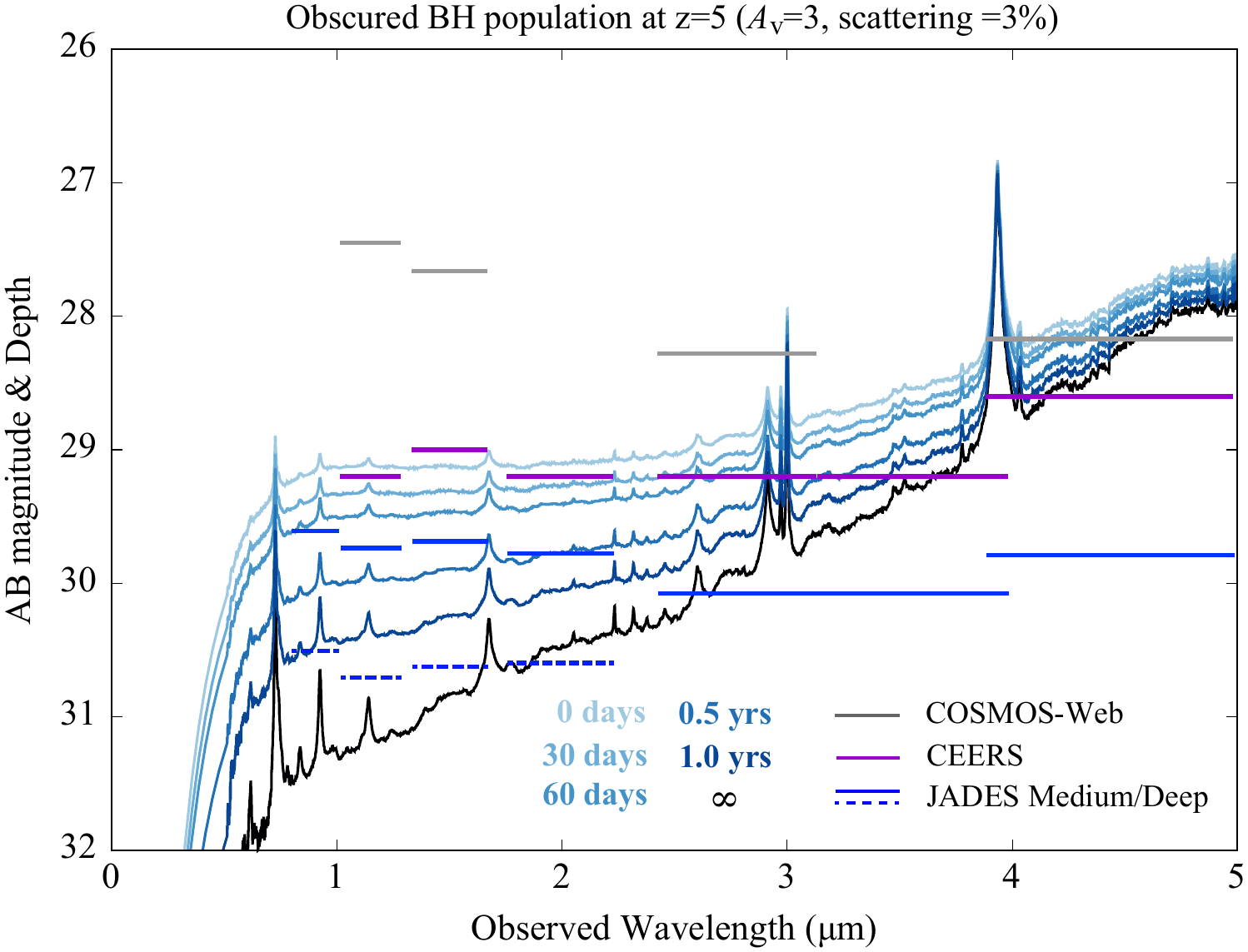}}
\caption{SEDs of stellar tidal disruption occurring in a dust-reddened, obscured AGN with $M_\bullet =10^6~\msun$ at $z=4$ (left) and $z=5$ (right), respectively. 
Dust extinction with $A_{\rm V}=3$ mag and a scattering fraction of $f_{\rm scatt}=3\%$ are assumed both for the TDE and AGN components 
(see Eq.~\ref{eq:SED}). 
Each curve indicates the time series of the SEDs since the peak time in the observer frame; $t-t_{\rm peak}=0$ days, $30$ days, $60$ days, $0.5$ years, 
$1.0$ years, and $\infty$ (only the AGN spectrum) from the top to the bottom.
We overlay the $5\sigma$ point-source imaging depths in each NIRCam filter of JWST survey programs; COSMOS-Web (grey), 
CEERS (purple), and JADES Medium/Deep (blue solid and dashed). 
Note that we show the depths of only four short-wavelength filters in JADES Deep for illustrative purposes.
}
\label{fig:SED_z45}
\end{center}
\end{figure*}

For a wide range of the parameters, the peak accretion rate exceeds the Eddington value significantly and reaches 
$\dot{M}_{\rm fb}/\dot{M}_{\rm Edd}\simeq O(10-100)$.
Despite a large gas supplying rate to the nuclear scale, powerful outflows are launched from the vicinity of the BH horizon
and carry a large fraction of the inflowing mass and momentum outward 
\citep[e.g.,][]{Ohsuga_2009,Jiang_2014,McKinney_2015,Sadowski_2015,Hu_2022a}.
As a result, the inflow rate decreases toward the center as $\dot{M}_{\rm in}(r)\propto r^q$ ($q>0$) and thus the BH feeding rate is 
reduced to a mildly super-Eddington accretion rate of $\sim 2-10~\dot{M}_{\rm Edd}$ \citep[][$q\sim 1$]{Hu_2022a}.
The reduction factor is not well understood, either in theory or through observations.
However, it could potentially be determined by examining the ratio between the inflow radius (e.g., the radius where rapid 
circularization of stellar debris occurs following a disruption) and the radius of the inner-most stable circular orbit around the central BH.
Furthermore, the source of TDE emission in the optical-to-UV bands has been debated to be either reprocessing of the X-ray emission 
from the accretion disk \citep[e.g.,][]{Guillochon_2013,Roth_2016}, or emission from outer shocks between the debris streams 
when they collide \citep[e.g.,][]{Piran_2015}, or perhaps some combination of both \citep[see more references in][]{vanVelzen_2020}.

In this paper, we adopt the fitting result of TDE light curves in the optical-to-UV bands summarized in \citet{vanVelzen_2020},
instead of modeling the TDE light curves based on the theoretical frameworks.
This approach allows us to avoid numerous uncertainties in modeling of the BH feeding, mass loading into outflows, circularization 
of stellar debris, and optical-UV emission mechanisms.
Among 33 TDEs listed in Table~2 of \citet{vanVelzen_2020}, we consider the light curve fit for PS1-10jh, which was the first TDE detected 
with a well-sampled rise to peak in optical survey data \citep{Gezari_2012}.
The spectral shape is fitted by a diluted black body spectrum with an effective temperature of $T_0\simeq 10^{4.59}~\K$, and 
the bolometric luminosity is estimated as $L_{\rm peak}\simeq 10^{44.47}~{\rm erg~s}^{-1}$ at the peak and subsequently decays following 
a $\propto t^{-5/3}$ power-law.
\citet{vanVelzen_2019} estimate the mass of the SMBH that causes the TDE PS1-10jh using the $M_\bullet - \sigma$ relation 
as $M_\bullet \simeq 10^{6.06}~\msun$.
They also find that the light curves of TDEs from such low-mass BHs ($M_\bullet < 10^{6.5}~\msun$) show 
significant late-time flattening at several years after the peaks.
The observed late-time emission is consistently described by a viscous-driven accretion disk model (i.e., AGN emission).

Motivated by those observational facts, we model the TDE light curve with a power-law decay and a plateau component as
\begin{align}
L_{\nu}(t) &= L_{\rm peak}\left(\frac{t+t_0}{t_0}\right)^{p}\frac{\pi B_\nu(T_0)}{\sigma_{\rm SB} T_0^4}
+ \frac{L_{\rm UV,\bullet}}{\nu_0}f_\nu,
\label{eq:SED}
\end{align}
where $\nu_0=c/\lambda_0=2\times 10^{15}~{\rm Hz}$, $t_0$ is the characteristic decay timescale,
$p(=-5/3)$ is the power-law index of the light-curve decay, $B_\nu(T_0)$ is the Planck function, $\sigma_{\rm SB}$ is the Stefan-Boltzmann constant.
The UV luminosity of the AGN, as a steady source, is estimated as $L_{\rm UV,\bullet} = \lambda_{\bullet}L_{\rm Edd}/f^{\rm bol}_{1500}$,
where $\lambda_{\bullet}\simeq 0.3-1.0$ is the Eddington ratio for the steady AGN emission and $f^{\rm bol}_{1500}\simeq 4.4$ is 
the bolometric correction factor for the monochromatic UV band.
The AGN spectral shape $f_\nu$ is taken from the composite spectrum of low-redshift quasars (see \citealt{VandenBerk_2001})
and its normalization is set so that $f_{\nu_0}=1$.
We take into account dust attenuation by the extinction law of starburst galaxies \citep{Calzetti_2000} and apply it both to the AGN and TDE spectra
in the form of $L_\nu e^{-\tau_\nu}$, where $\tau_\nu$ is calculated from the extinction law. 
For unobscured systems, we fix the level of dust extinction to $A_{\rm V}\simeq 0.3$ mag, the average value of extinction for JWST-detected 
broad-line AGNs with $M_\bullet<10^7~\msun$ presented in Table~3 of \cite{Maiolino_2023_JADES}.
On the other hand, we consider a higher level of extinction with $A_{\rm V}=3.0$ mag for obscured systems, 
motivated by the redden rest-optical continuum spectra seen in dust-reddened, obscured broad-line AGNs \citep{Kocevski_2023,Matthee_2023,Labbe_2023}.
The observed spectra of those red sources also show the blue excess at shorter wavelengths of $\lambda_{\rm rest}<0.3-0.4~\mum$,
requiring additional blue components.
Although the origin of the blue excess remains uncertain yet, in the analogy of SEDs for the blue-excess dust obscured galaxies,
a small fraction $f_{\rm scatt}$ of the radiation flux of the intrinsic AGN spectrum would be scattered to our line of sight and develops the blue component
\citep{Zakamska_2005,Polletta_2006,Alexandroff_2018,Noboriguchi_2022,Noboriguchi_2023}.
In this paper, we adopt $f_{\rm scatt}=0.03$ to reproduce the spectral shapes of dust-reddened AGNs reported in the JWST UNCOVER field 
\citep[e.g.,][]{Greene_2023}, leading to the luminosity density of $L_\nu (f_{\rm scatt}+e^{-\tau_\nu})$.

\begin{table*}
\renewcommand\thetable{1} 
\caption{NIRCam imaging depths in various JWST survey programs.}
\begin{center}
\begin{tabular}{cccccc}
\hline
\hline
Filter & COSMOS-Web & CEERS & JADES-Medium & JADES-Deep &NGDEEP \\
\hline 
F090W  &  --       & --      & 29.60 & 30.50 & -- \\
F115W  &  27.45 & 29.20 & 29.73 & 30.70 & 31.20 \\
F150W  &  27.66 & 29.00 & 29.68 & 30.62 & 30.90 \\
F200W  &  --       &  29.20 & 29.77 & 30.59 & 30.90 \\
F277W  &  28.28 & 29.20  & 30.07 & 30.96 & 30.90\\
F356W  &  --       &  29.20  & 30.07 & 30.89 & 30.80\\
F410M &  --        & 28.40   & 29.62 & 30.52 & --\\
F444W &  28.17 &  28.60  & 29.78 & 30.64 & 30.70\\
\hline 
\end{tabular}
\label{tab:jwst_mag}
\end{center}
\tablecomments{~Column (1): NIRCam filters used for each survey. Column (2)-(6): average 5$\sigma$ point-source depths for 
COSMOS-Web \citep{Casey_2023}, CEERS \citep{Finkelstein_2023}, and JADES Medium/Deep \citep{Eisenstein_2023}, and 
NGDEEP \citep{Bagley_2023}.
}
\end{table*}

In our analysis, we do not consider the contribution of the underlying host galaxy to the observed SED for the following reason.
The rest-frame UV luminosity ($L_{\rm UV,\star}$) to host stellar mass ($M_\star$) relationship can be approximated as 
\begin{equation}
\frac{L_{\rm UV,\star}}{10^{43}~{\rm erg~s}^{-1}} \sim 1.7f_\star \left(\frac{M_\star}{10^9~\msun}\right)
\left(\frac{t_{\rm age}}{1~{\rm Gyr}}\right)^{-1},
\end{equation}
using the {\tt STARBURST99} population synthesis code (version 7.0.1; \citealt{Leitherer_1999}), 
with a Kroupa IMF (\citealt{Kroupa_2001}; $0.1-100~\msun$), Padova isochrone models, constant star formation, solar metallicity, and
the stellar age of $t_{\rm age}=1{\rm Gyr}$. 
While the value of $f_\star$ varies based on these assumptions, it typically remains $f_\star \sim O(1)$.
As a result, the ratio of UV luminosity between the AGN and host galaxy is expressed as
\begin{equation}
\frac{L_{\rm UV,\bullet}}{L_{\rm UV,\star}} \simeq 8.4 ~\left(\frac{\lambda_\bullet}{f_\star}\right) \left(\frac{M_\bullet/M_\star}{5\times 10^{-3}}\right).
\end{equation}
Therefore, as long as the $M_\bullet/M_\star$ ratio is consistent with the local values ($\sim 0.5\%$; \citealt{Kormendy_Ho_2013})
or above as suggested by JWST-identified AGNs (see also \citealt{Pacucci_2023}), the UV luminosity from the AGN dominates over 
the host galaxy contribution, assuming similar levels of extinction for both.

\begin{figure*}
\begin{center}
{\includegraphics[width=88mm]{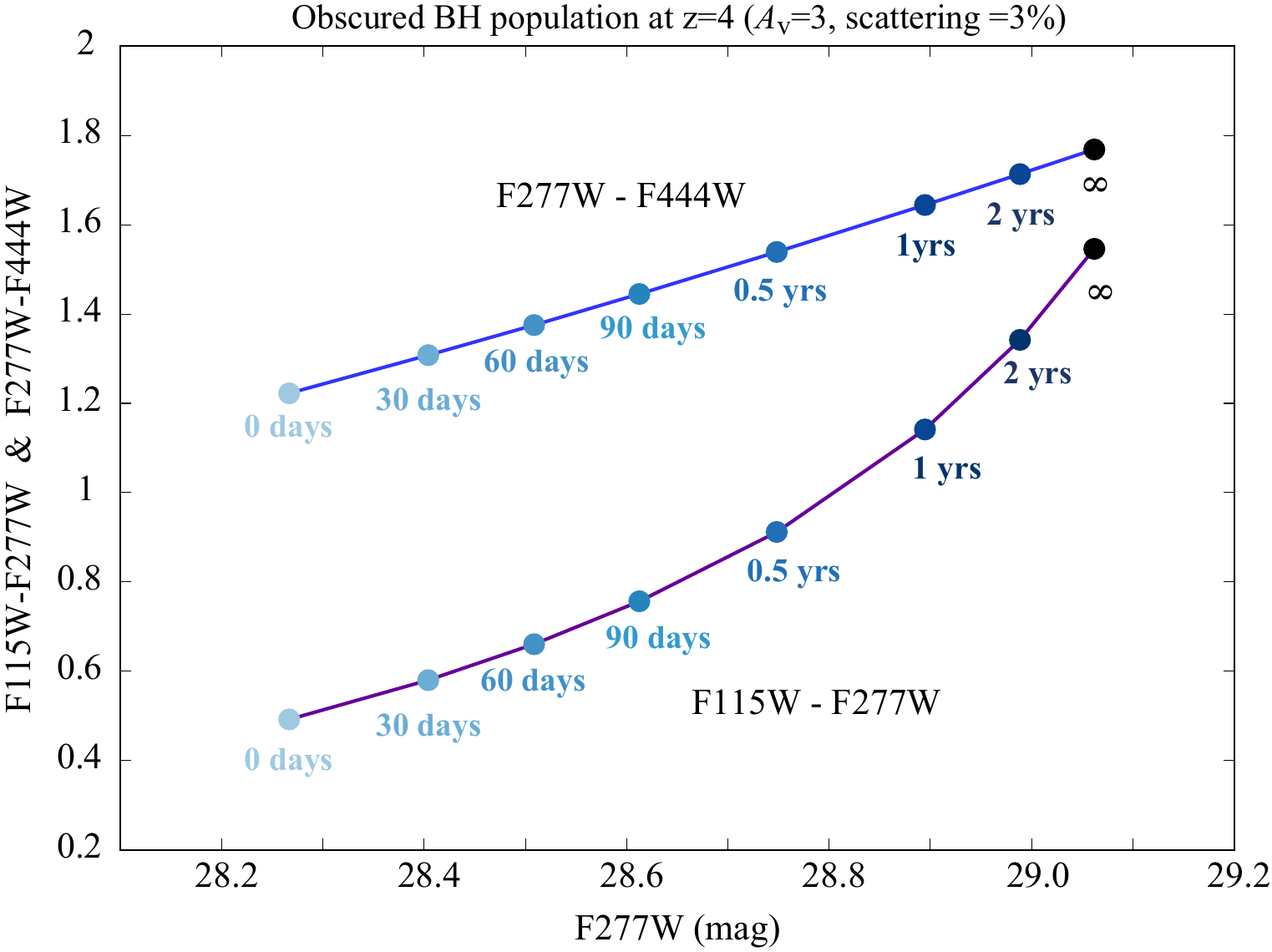}}\hspace{2mm}
{\includegraphics[width=88mm]{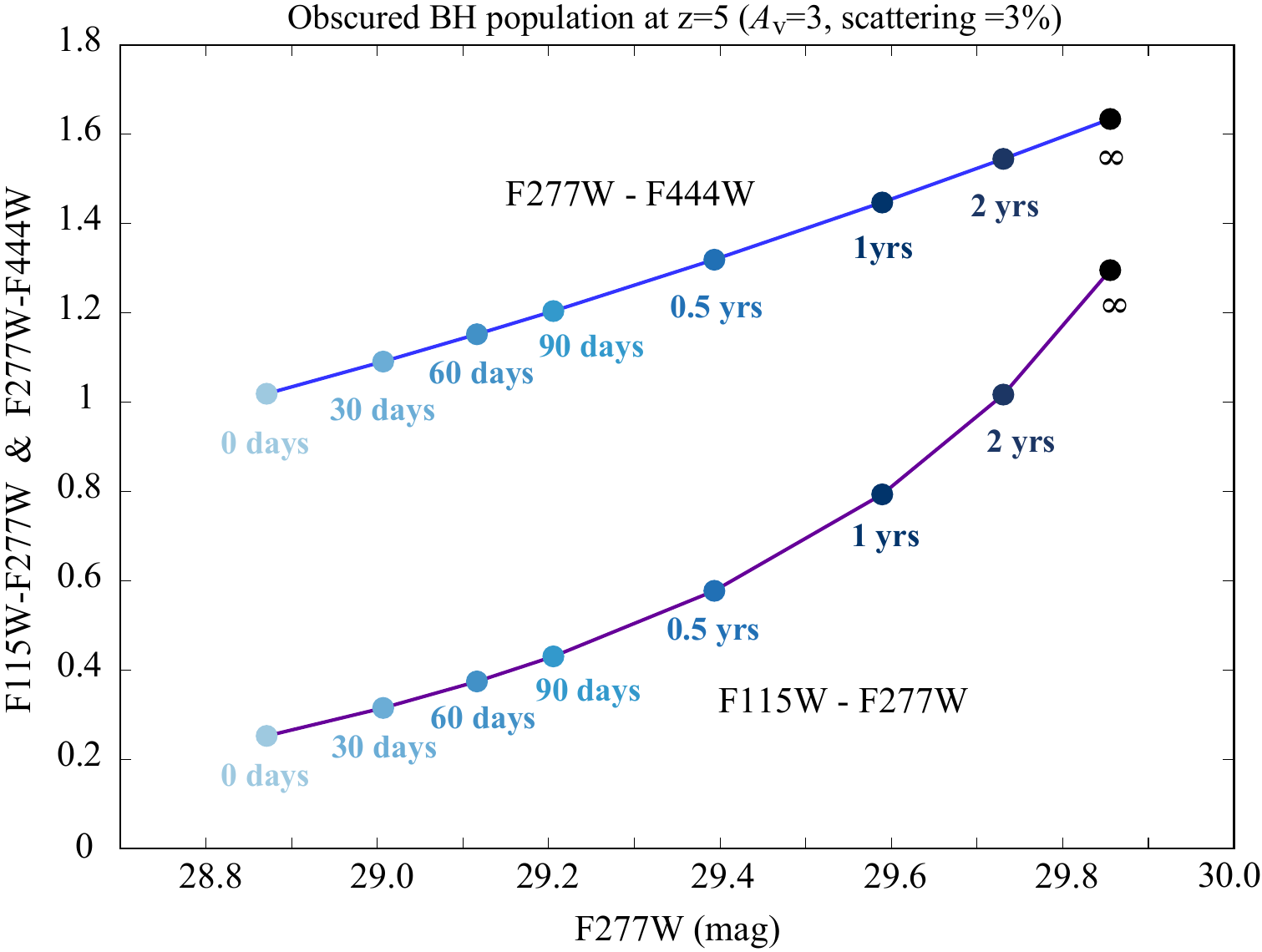}}
\caption{The color-magnitude diagram for TDEs in a red obscured AGN with $M_\bullet =10^6~\msun$ at $z=4$ (left) and $z=5$ (right), respectively.
Each curve presents the F277W--F444W (blue) and F115W--F277W colors (purple), respectively, and circle symbols indicate
the elapsed time since the TDE brightness peak.
}
\label{fig:color_z45}
\end{center}
\end{figure*}

\subsection{Deep-field surveys for obscured TDEs}
\label{sec:deepobs}

We first consider stellar TDEs occurring within dust-reddened, obscured AGNs.
In this scenario, we assume dust extinction of $A_{\rm V}=3$ mag and a scattering fraction of $f_{\rm scatt}=0.03$ 
for both the TDE and AGN components, as described in Eq.~(\ref{eq:SED}).
Due to dust extinction, the rest-frame UV light on the total SED is significantly attenuated.
Consequently, deep surveys with JWST are required to meet the detection threshold.
However, as a trade-off for achieving such depths in observations, the survey area becomes severely limited ($<0.6~{\rm deg}^2$), given the limited amount of observing times.
Therefore, we need to consider a more abundant population of BHs providing TDEs in an efficient way; specifically those 
with low masses with $M_\bullet \lesssim 10^6~\msun$ (see the solid curves in Figure~\ref{fig:tde_rate}).

In Figure~\ref{fig:SED_z45}, we present the SEDs of stellar tidal disruption in an obscured AGN with $M_\bullet =10^6~\msun$
at $z=4$ (left) and $z=5$ (right), respectively.
Each curve indicates the time series of the SED at each epoch since the TDE peak in the observer frame; $t-t_{\rm peak}=0$ days, $30$ days, $60$ days, $0.5$ years, 
$1.0$ years, and $\infty$ (TDE emission is neglected) from the top to the bottom.
We overlay the $5\sigma$ point-source imaging depths in each NIRCam filter used for JWST survey programs; COSMOS-Web, 
CEERS, and JADES Medium/Deep (see also the summary of the depths for each filter in Table~\ref{tab:jwst_mag}).
In Figure~\ref{fig:color_z45}, we show the color-magnitude diagram for this obscured TDE+AGN at $z=4$ (left) and $z=5$ (right).
Each curve presents the F277W--F444W (blue) and F115W--F277W colors (purple), respectively, and the circle symbols indicate
the elapsed time since the peak brightness of the TDEs.

At the lower redshift ($z=4$), the SED enables the robust detection of the red continuum component using the longer-wavelength filters 
(F277W, F356W, and F444W) in the three surveys: COSMOS-Web, CEERS, and JADES.
As the TDE light curve decays over time, the F277W flux density declines by $\sim 1$ mag in one year, resulting in a reddening of the F277W--F444W color.  
Furthermore, the scattered light from the TDE introduces a blue excess component (i.e., rest-frame UV bands), which is observable 
within $1-2$ months since the luminosity peak in the CEERS and JADES surveys. 
The dominance of TDE-originating light on the blue side of the spectrum allows for the direct tracer of 
the $t^{-5/3}$ decay rate using the F115W--F277W color in the JADES-Deep survey.

At the higher redshift ($z=5$), the overall trend of the SED shape still holds. 
However, due to the far distance, the F277W flux density becomes too low for detection in the CEERS and JADES-Medium survey sensitivities
after $\sim 0.5$ years from the TDE peak.
The blue excess in the spectrum can be only detectable with the JADES-Deep survey until the same elapsed time,
when the F115W--F277W color reaches $\simeq 0.6$ mag.
In the subsequent epochs, the TDE is identified only with longer-wavelength filters and the F277W--F444W color becomes redder with time.

\begin{figure*}
\begin{center}
{\includegraphics[width=88mm]{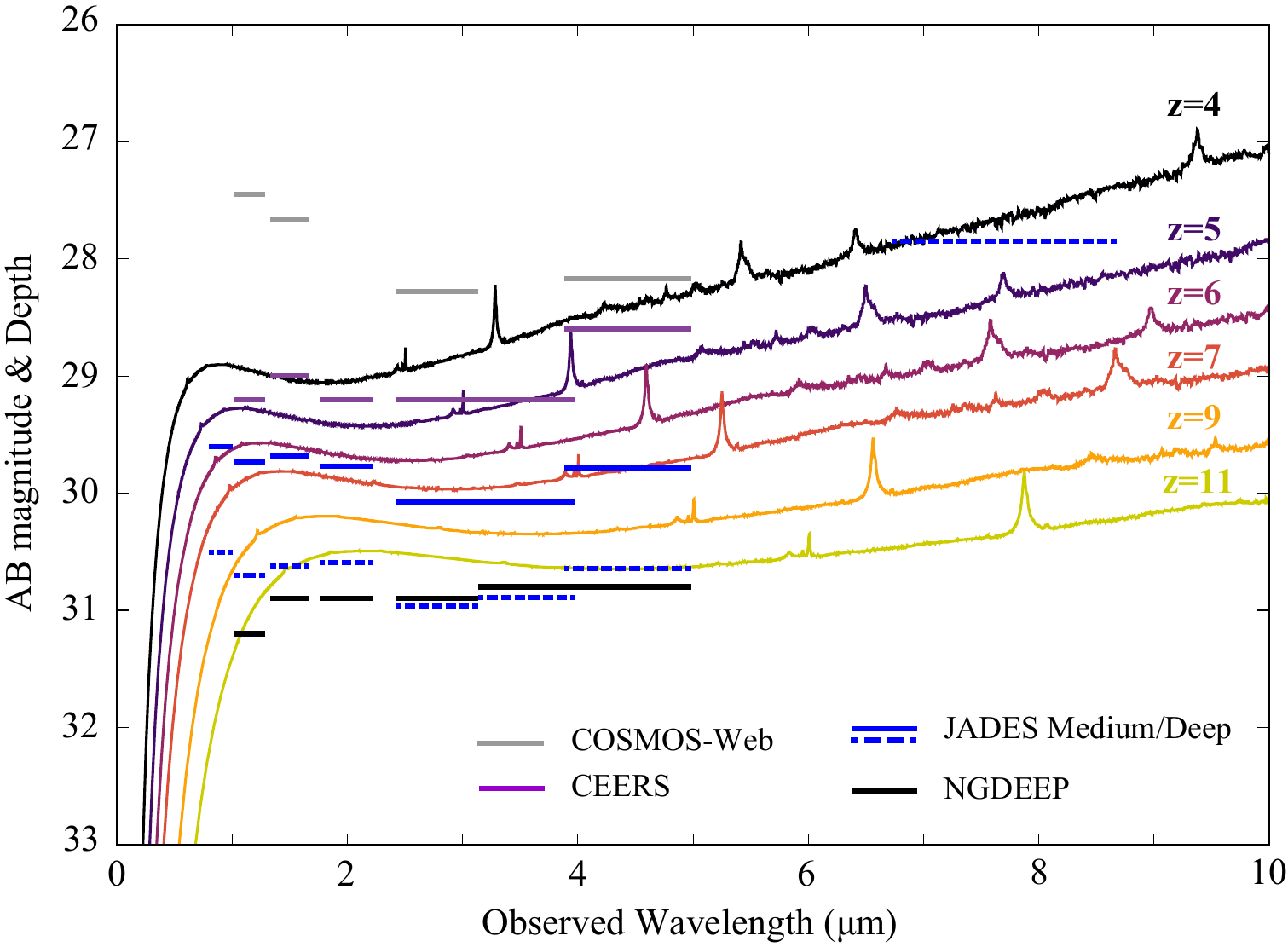}}\hspace{2mm}
{\includegraphics[width=88mm]{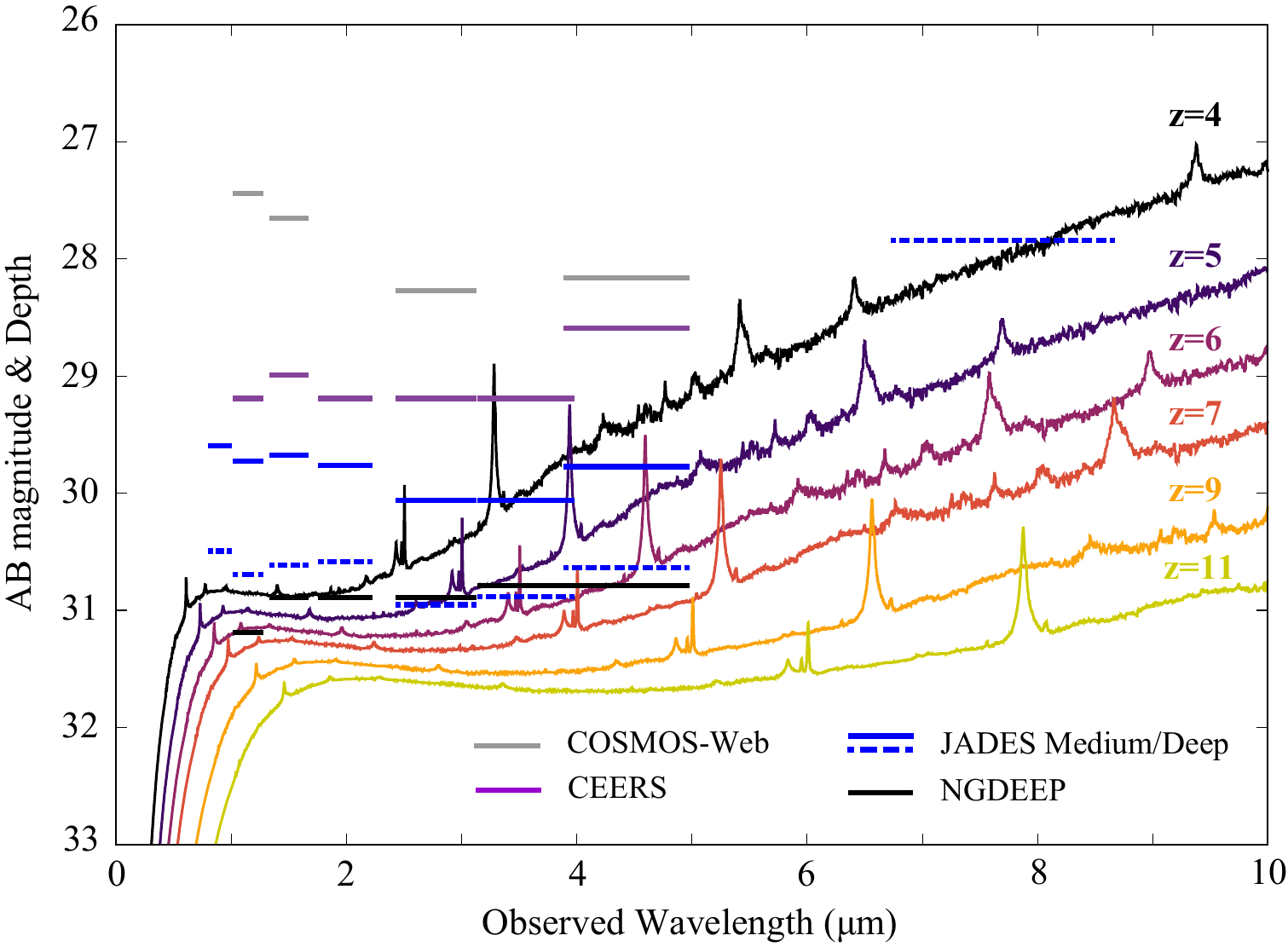}}
\caption{SEDs of stellar tidal disruption occurring in a dust-reddened, obscured AGN with $M_\bullet =10^5~\msun$ over $z=4-11$, 
at the elapsed time of $t-t_{\rm peak}=0$ days (left) and 1 year (right) since the TDE peak time, respectively. 
We overlay the $5\sigma$ point-source imaging depths in each NIRCam filter of JWST survey programs; COSMOS-Web (grey), 
CEERS (purple), JADES Medium/Deep (blue solid and dashed), and NGDEEP (black).
}
\label{fig:SED_highz}
\end{center}
\end{figure*}

\begin{figure}
\begin{center}
{\includegraphics[width=85mm]{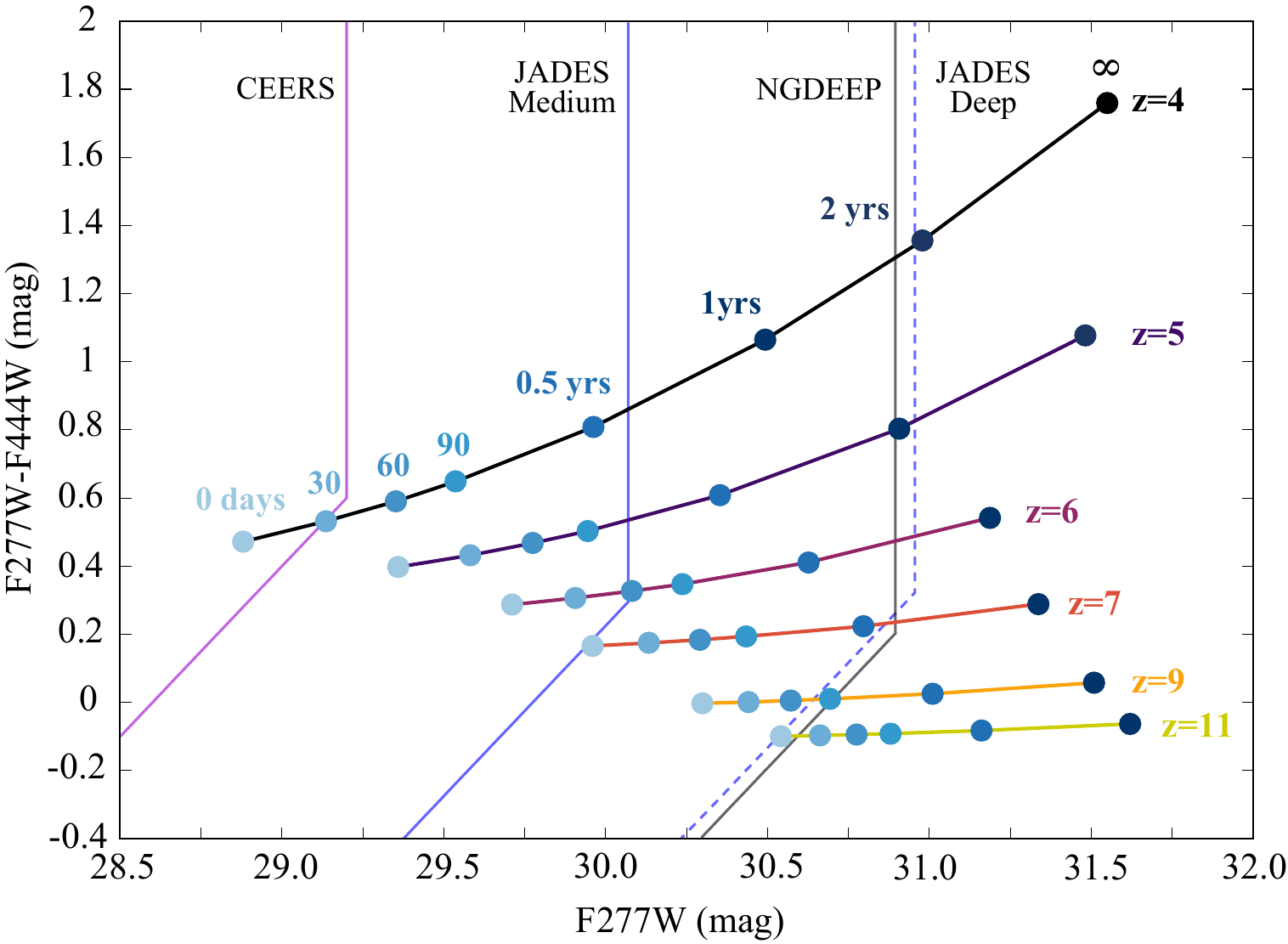}}
\caption{The color-magnitude diagram for TDEs in a red obscured AGN with $M_\bullet =10^5~\msun$ over $z=4-11$.
Each curve presents the F227W--F444W (blue) and F115W--F277W colors (purple), respectively, and circle symbols indicate
the elapsed time since the TDE brightness peak.
}
\label{fig:color_highz}
\end{center}
\end{figure}

Toward higher redshifts ($z>5$), only deep surveys such as JADES/Deep \citep{Eisenstein_2023} and NGDEEP \citep{Bagley_2023}, 
which achieve limiting magnitudes of $\sim 31$ mag but cover smaller areas of $<0.01~{\rm deg}^2$.
In this scenario, we consider a further abundant population of BHs with $M_\bullet =10^5~\msun$.
Even within such limited fields of view, we expect to detect $\sim O(1)$ TDEs triggered by the low-mass BH.
Figure~\ref{fig:SED_highz} illustrates the SEDs of stellar tidal disruption occurring in a dust-reddened, obscured AGN with $M_\bullet =10^5~\msun$ 
at various redshifts from $z=4$ to $11$.
Each panel represents the SED shapes at two different time points: immediately after the TDE peak time (left) and one year after the peak (right).
Figure~\ref{fig:color_highz} shows the color-magnitude diagram for the TDE with longer-wavelength NIRCam filters.
The JADES-Medium survey is capable of identifying the characteristic colors of the TDE in the early stages at $z\lesssim 7$, 
while the detection horizon extends to as far as $z\sim 11$ in the JADES/Deep and NGDEEP surveys.
Even within one year since the TDE peak, the deepest two surveys capture the red continuum component at wavelengths around $\sim 3-5~\mum$.

\subsection{Wide-field surveys for unobscured TDEs}
\label{sec:wideobs}

Next, we consider TDEs occurring within unobscured AGNs.
In this scenario, we adopt dust extinction of $A_{\rm V}=0.3$ mag for both the TDE and AGN components.
While this or low level of extinction exerts negligible influence on the total SED, the intrinsic TDE rate is approximately two orders of magnitude lower than 
that for the obscured cases (see Section~\ref{sec:deepobs}).
As a result, the detection of unobscured TDEs requires wide-field surveys with RST and the COSMOS-Web survey program with JWST.
Specifically, to detect more than one TDE at $z\gtrsim 4$, the planned surveys need to target AGN populations with BH masses with 
$M_\bullet \gtrsim 10^5~\msun$, $10^6~\msun$, and $3\times 10^7~\msun$ for survey areas larger than $\sim 2.5~{\rm deg}^2$, $50~{\rm deg}^2$, 
and $1,700~{\rm deg}^2$, respectively (see Figure~\ref{fig:tde_rate}).

\begin{figure*}
\begin{center}
{\includegraphics[width=88mm]{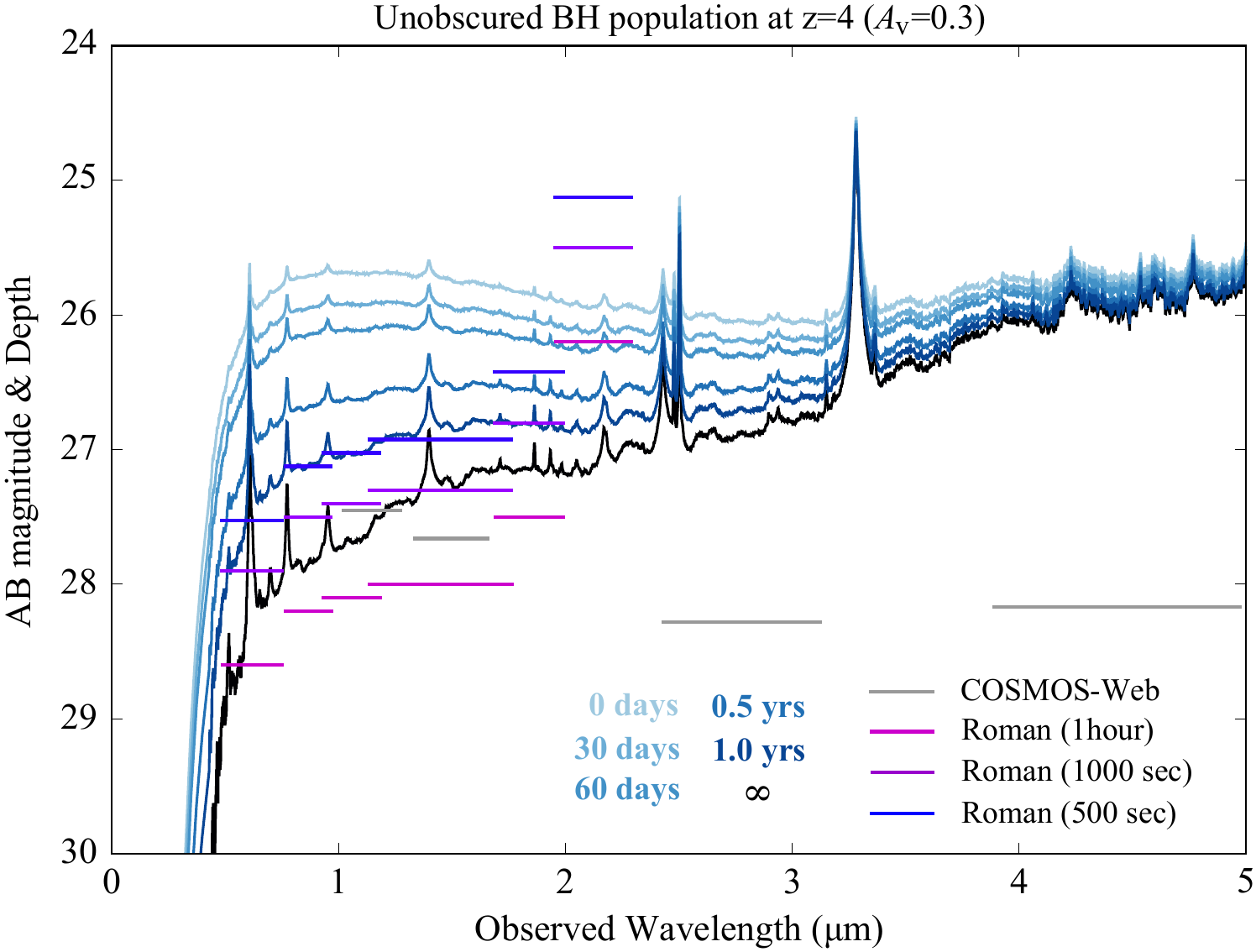}}\hspace{2mm}
{\includegraphics[width=88mm]{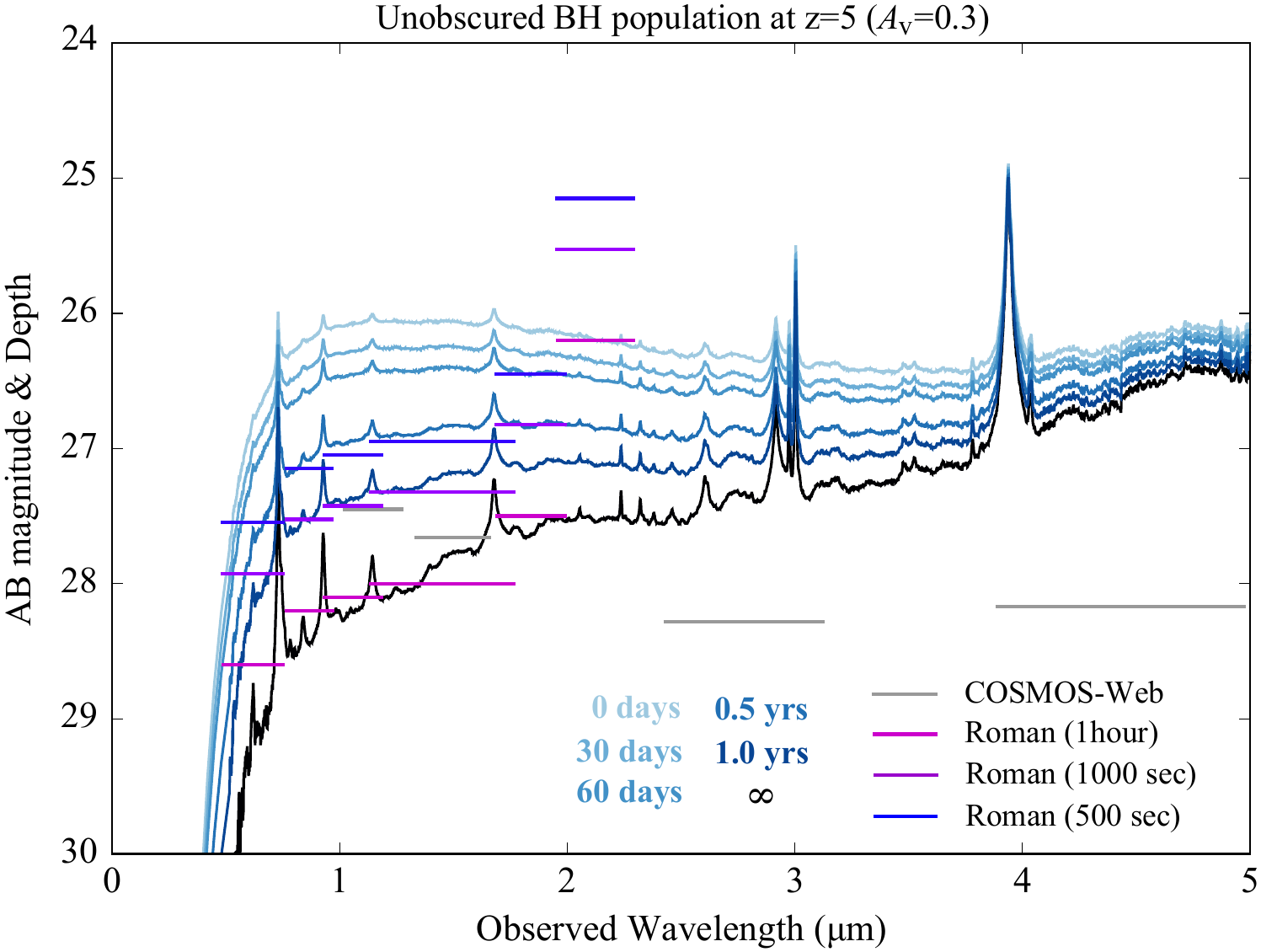}}\vspace{2mm}\\
{\includegraphics[width=88mm]{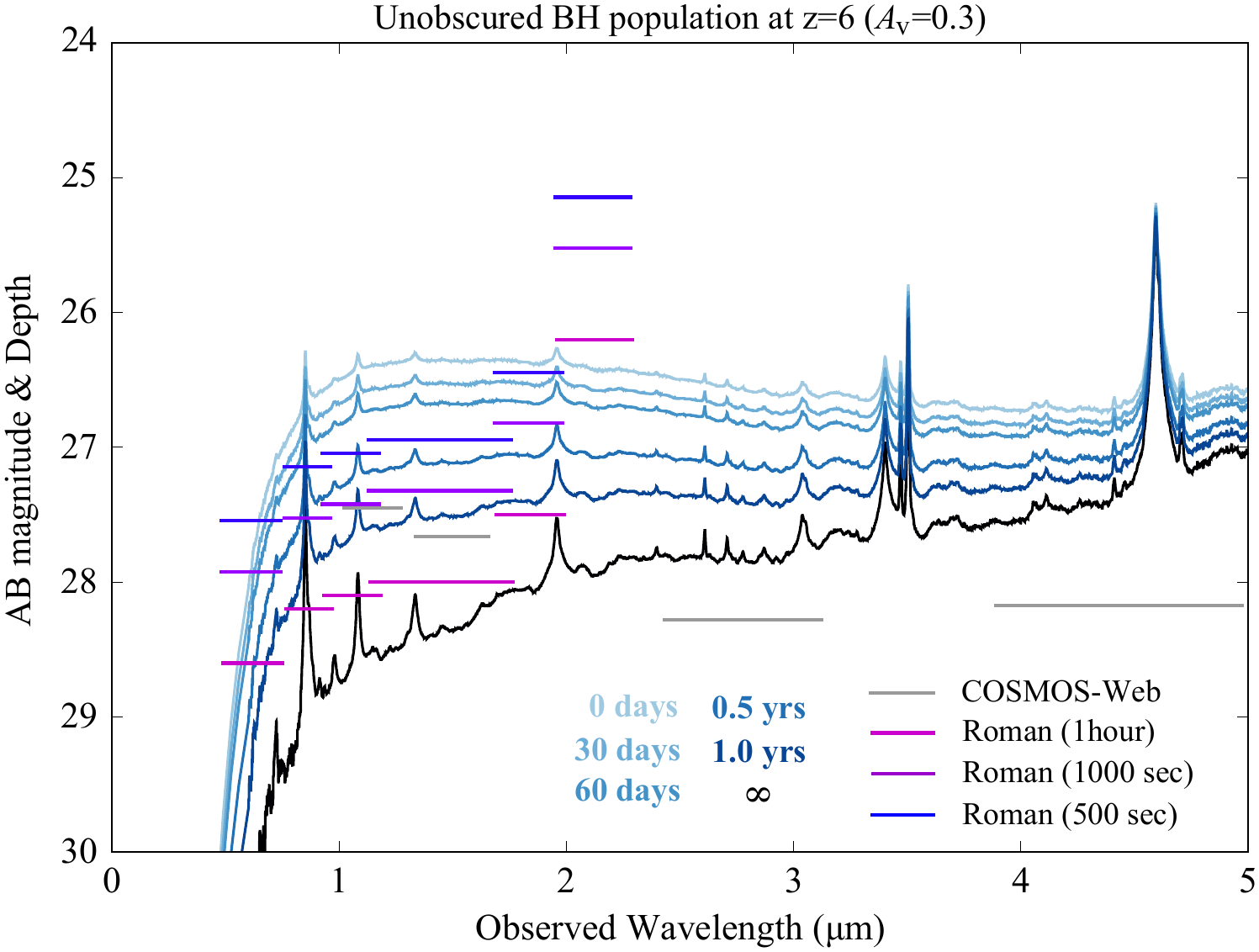}}\hspace{2mm}
{\includegraphics[width=88mm]{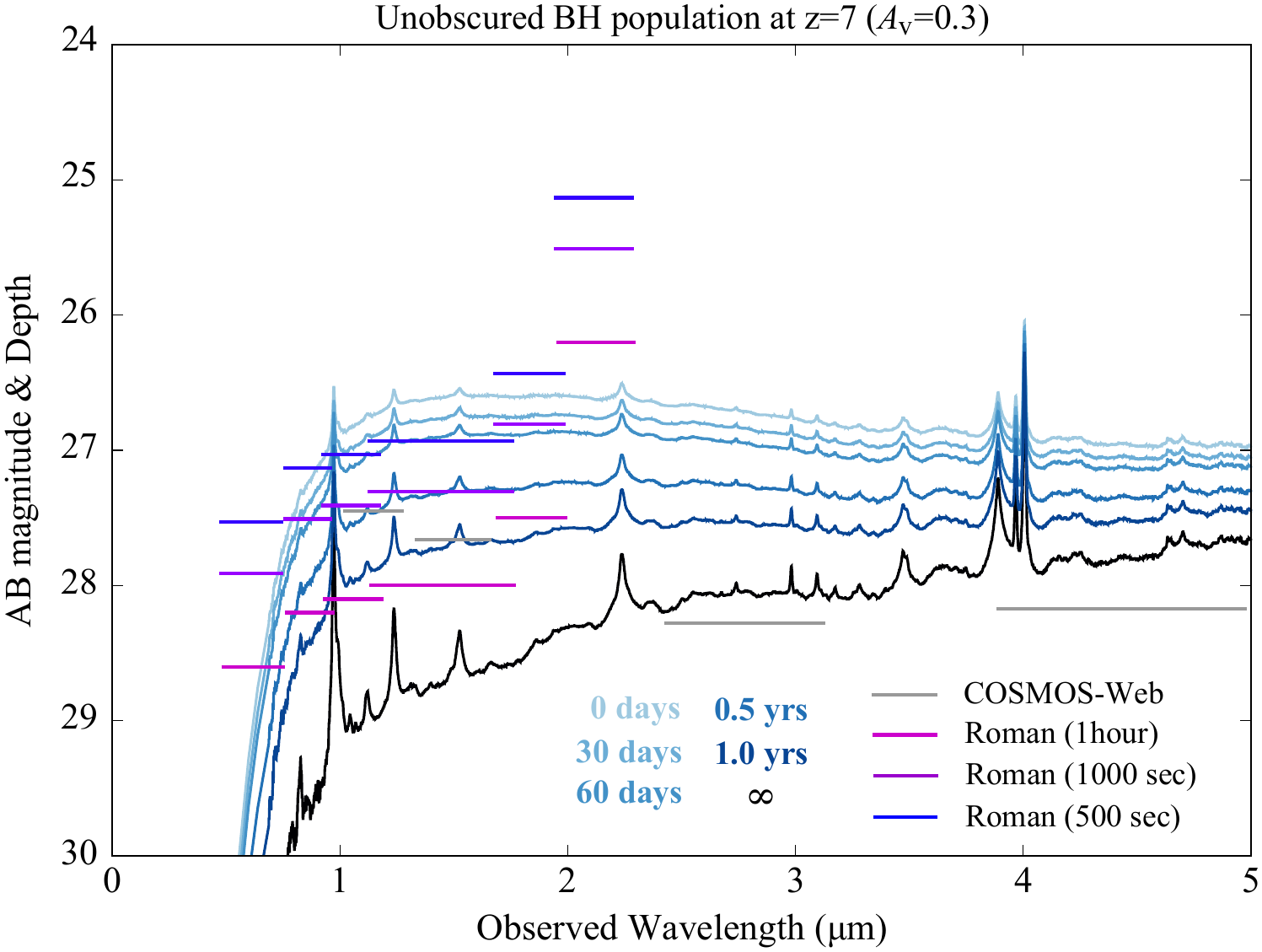}}\\
\caption{SEDs of stellar tidal disruption occurring in an unobscured AGN with $M_\bullet =10^6~\msun$ at $z=4$ (top left), $z=5$ (top right),
$z=6$ (bottom left), and $z=7$ (bottom right), respectively. 
Dust extinction with $A_{\rm V}=0.3$ mag is assumed both for the TDE and AGN components (see Eq.~\ref{eq:SED}). 
Each curve indicates the time series of the SEDs since the peak time; $t-t_{\rm peak}=0$ days, $30$ days, $60$ days, $0.5$ years, 
$1.0$ years, and $\infty$ (only the AGN spectrum) from the top to the bottom.
We overlay the $5\sigma$ point-source imaging depths in each filter of RST and JWST (COSMOS-Web) observations.
For the RST observations, we consider 1 hour, $10^3$ second , and $500$ second of exposure time.}
\label{fig:SED_highz_unobs}
\end{center}
\end{figure*}

In Figure~\ref{fig:SED_highz_unobs}, we present the SEDs of stellar tidal disruption in an unobscured AGN with $M_\bullet =10^6~\msun$
at various redshifts from $z=4$ to $z=7$, respectively.
Each curve indicates the time series of the SED at each epoch since the TDE peak; $t-t_{\rm peak}=0$ days, $30$ days, $60$ days, $0.5$ years, 
$1.0$ years, and $\infty$ (TDE emission is neglected) from the top to the bottom.
To assess the detectability of these events, we overlay the $5\sigma$ point-source imaging depths in each filter of RST and JWST (COSMOS-Web) observations.
For the RST observations, we consider single-epoch exposure times of 1 hour (= 3.6 ks), $1.0$ ks , and $0.5$ ks\footnote{
\url{https://roman.gsfc.nasa.gov/science/Roman_Reference_Information.html}}.
In the cases of unobscured TDEs, the observed flux at $<2~\mum$ is as bright as $\lesssim 26-27$ mag at the peak time depending on the redshifts
and decays within one year down to $\lesssim 27-28$, which remains detectable with RST in an exposure time of $\sim 1$ ks.
Similar observations can be conducted using JWST in the COSMOS-Web survey, which offers additional advantage of covering the TDE spectrum with two longer-wavelength filters,
F277W and F444W, and monitoring the characteristic decay of the TDE emission.

\begin{table}
\renewcommand\thetable{2} 
\caption{Single-exposure depths of future wide-field surveys.}
\begin{center}
\begin{tabular}{cc|cc|cc}
\hline
\hline
Filter & RST Wide  & Filter & Euclid & Filter & LSST\\
\hline 
  &          & & & $u$ &23.9 \\
  &          & & & $g$ & 25.0\\
F062 &  27.1   & \multirow{3}{*}{$I_{\rm E}$} &\multirow{3}{*}{25.3} & $r$ & 24.7 \\
\multirow{2}{*}{F087}  & \multirow{2}{*}{26.3}  & & & $i$ & 24.0  \\
  &   & & & $z$ & 23.3\\
F106  &  26.2 & $Y_{\rm E}$ & 24.0 & $y$ & 22.1\\
F129  &  26.1 & $J_{\rm E}$ & 24.0 &        & \\
          &          & $H_{\rm E}$ & 24.0 &       & \\
\hline 
\end{tabular}
\label{tab:rst_mag}
\end{center}
\tablecomments{~Average $5\sigma$ point-source depths for each filter in the RST wide tier of the High-latitude Time Domain Survey 
for supernova cosmology \citep{Rose_2021}, Euclid \citep{Laureijs_2011},
and LSST \citep{Bianco_2022}.
}
\end{table}

\begin{figure}
\begin{center}
{\includegraphics[width=85mm]{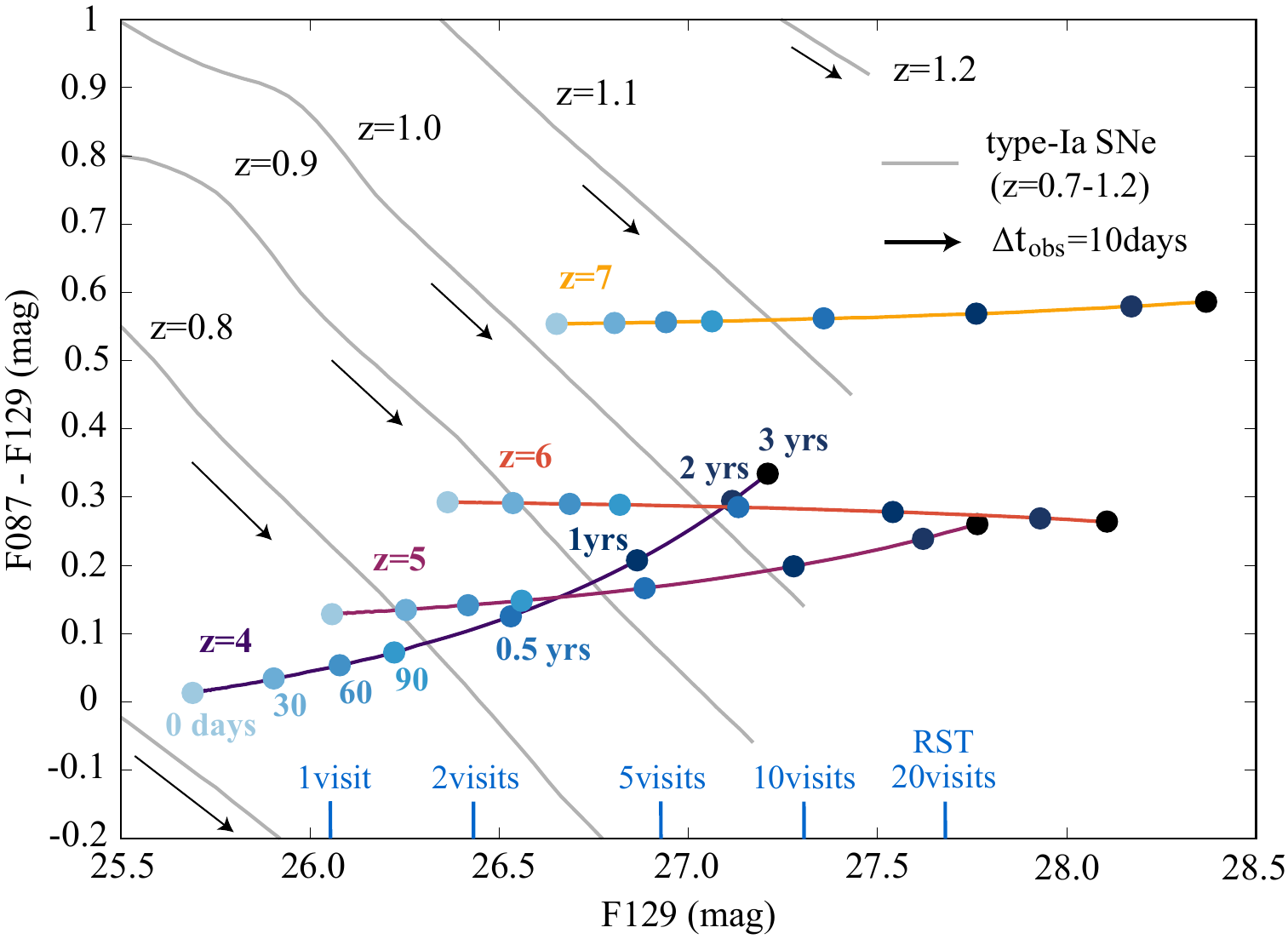}}
\caption{The color-magnitude diagram for TDEs in an unobscured AGN with $M_\bullet =10^6~\msun$ over $z=4-7$.
Each curve presents the F087--F129 colors as a function of the F129 flux density assuming the RST filter transmissions, and circle symbols indicate the elapsed time since the TDE brightness peak.
The grey curves are the colors and magnitudes of type-Ia SNe at $z=0.7-1.2$ along with black arrows that indicate the direction of the time evolution
and the length corresponds to a 10-days duration in the observed frame.
The limiting magnitude of the F129 filter for multiple exposures ($N=1$, $2$, $5$, $10$, and $20$), where a single exposure time is assumed to be 100 sec.
}
\label{fig:color_highz_unobs}
\end{center}
\end{figure}

Figure~\ref{fig:color_highz_unobs} shows a color-magnitude diagram for TDEs in an unobscured AGN with $M_\bullet =10^6~\msun$ over $z=4-7$.
Each curve presents the F087--F129 colors as a function of the F129 flux density assuming the RST filter transmissions.
Circle symbols indicate the elapsed time since the peak brightness of the TDEs.
This particular filter combination, F087 and F129, is chosen in alignment with the initial survey design for the High-latitude Time Domain Survey 
for supernova cosmology with RST \citep[e.g.,][]{Rose_2021}.
This survey employs a two-tier strategy: wide and deep.
The wide tier aims to cover the RZYJ bands (F062, F087, F106, and F129) across a survey area of $\simeq 20~{\rm deg}^2$,
while the deep tier focuses on the YJHF bands (F106, F129, F158, and F184) with a survey area of $\simeq 4~{\rm deg}^2$.
The entire survey spans 6 month observing time spread over 2 years, with 30-hour visits occurring every 5 days.
The expected $5\sigma$ point-source imaging depths are 26.3 mag in the F087 filter and 26.1 mag in the F129 filter for 
each single-epoch exposure with 100 sec, as summarized in Table~\ref{tab:rst_mag}.
Therefore, to capture the early-stage TDE emission ($\sim 90$ days in the observed frame, equivalent to 18 visits), 
it is necessary to cover 2-3 visits for TDEs at $z=4-5$ and at least 5-7 visits for TDEs at $z=6-7$.
This allows for the construction of a robust light curve with 3-9 data points during each stacking period.

In practice, there still remains potential contamination from nearby and lower-redshift astrophysical objects.
The High-latitude Time Domain Survey is primarily targeting type-Ia SNe at redshifts of $z>0.5$ as a key component of transient survey programs
aimed at measuring the expansion history of the universe.
The reference survey described in \citet{Rose_2021} is designed so that a substantial number, the order of $10^4$, of type-Ia SNe with 
a high signal-to-noise ratio (S/N$>10$) will be detected in the survey.
In Figure~\ref{fig:color_highz_unobs}, we also overlay the colors and magnitudes of type-Ia SNe at lower redshifts, specifically in the range of $z=0.7-1.2$ (grey curves).
We adopt the type-Ia SN spectral templates that span the time between peak luminosity and 85 days after the peak in the rest frame,
based on the work of \citet{Hsiao_2007}. 
The template luminosity is set to $-19.3$ mag at the peak in the $R$-band, which is the average of the standard type Ia SNe \citep{Yasuda_Fukugita_2010}.
The black arrows indicate the direction of time evolution for the colors and magnitudes, with the length of the arrows corresponding to a 10-days duration in the observed frame.
As clearly demonstrated, the colors and magnitudes of type-Ia SNe at $z\simeq 0.7-1.0$ in the later stages exhibit similar values to those of TDEs at $z\simeq 4-6$ in the earlier stages.
This similarity makes it challenging to distinguish high-$z$ TDEs from low-$z$ type-Ia SNe with only one or two exposures.
However, it is worth noting that the colors and magnitudes of these objects evolve by $\sim 0.1-0.2$ mag within a 10-day duration (equivalent to 3 visits) in the observed frame 
as indicated with the black arrows.
Importantly, the direction of the color-magnitude evolution in this diagram is nearly perpendicular to that of high-$z$ TDEs.
This distinct evolution pattern plays an essential role in identifying high-$z$ TDEs among lower-$z$ type-Ia SNe.
In addition to the color and magnitude criteria, Lyman dropout selection with the F062 and F087 bands can be used to distinguish $z>5.3$ transients
from those low-$z$ type-Ia SNe.

Other types of energetic explosions occurring at the high-redshift universe are also intriguing targets for the RST transient survey.
\citet{Moriya_2022} have explored the detectability of superluminous supernovae (SLSNe) or pair-instability supernovae (PISNe) with luminosities
comparable to the brightest TDEs.
Their analysis indicates that a significant number of SLSNe and PISNe at $5<z<6$ can be detectable and identified using the F158 and F213 filters\footnote{
The filter combination is motivated by the fact (and hypothesis) that both SLSNe and PISNe have SEDs peaked at $\lambda_{\rm rest} \simeq 3000~{\rm \AA}$,
corresponding to $\lambda_{\rm obs} \simeq 1.8~\mum$ at $z\sim 5$ (see more details in \citealt{Moriya_2022}).}, 
covering longer wavelengths than those considered for TDE searches in this paper.
Our filter combination with F087 and F129 can also be employed to detect these energetic supernovae at lower redshifts ($z\sim 2-3$). 
However, it is crucial to note that these explosions should be substantially brighter ($\lesssim 24.5$ mag) than TDEs.

\begin{figure}
\begin{center}
{\includegraphics[width=85mm]{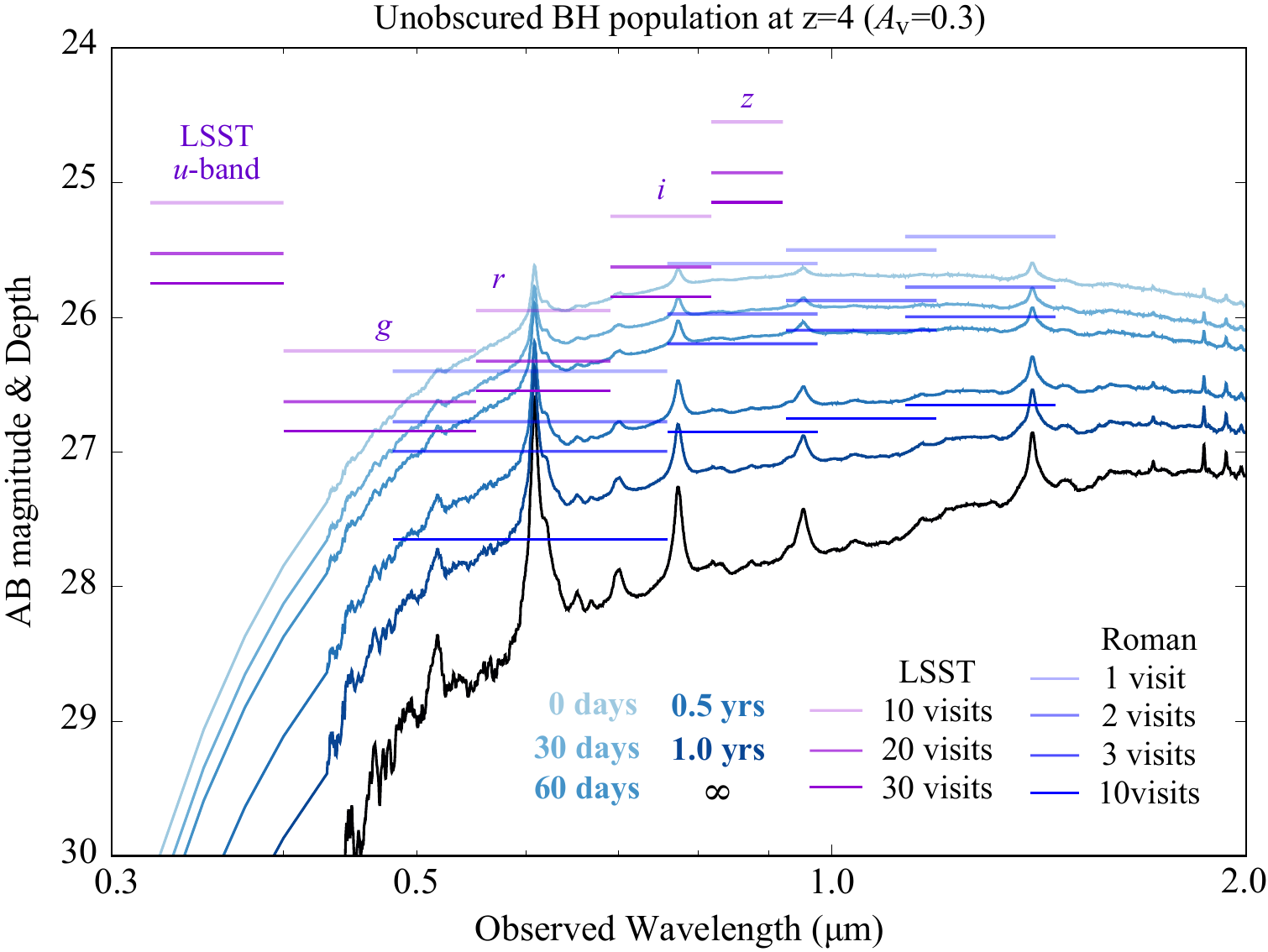}}
\caption{Synergy between the LSST and Roman wide survey for detecting a TDE occurring in an unobscured AGN ($A_V=0.3$ mag) with $M_\bullet = 10^6~\msun$ at $z = 4$.
We overlay the $5\sigma$ point-source imaging depths in each filter of LSST \citep{Bianco_2022} and RST (the wide survey; \citealt{Rose_2021}) for multiple exposures.
}
\label{fig:LSST_RST}
\end{center}
\end{figure}

\subsection{Other surveys for TDEs at $z\gtrsim 4$}
\label{sec:lsst_euclid}

We briefly discuss the capabilities of other wide-field surveys, such as LSST and Euclid, for the exploration of high-$z$ TDEs. 
The expected $5\sigma$ point-source depths for the LSST and Euclid bands are summarized in Table~\ref{tab:rst_mag}.

LSST is designed for high-cadence optical transient exploration, particularly suitable for detecting TDEs (or their candidates) at high redshifts. 
In Figure~\ref{fig:LSST_RST}, we present SEDs of a TDE occurring in an unobscured AGN with $M_\bullet=10^6~\msun$ at $z\sim 4$.
The blue part of the spectra at $\lambda_{\rm obs}\lesssim 0.7~\mum $ can be detectable through image stacking of $10-20$ visits in the $g$ and $r$ bands
up to $\sim 60$ days (in the observer frame) after the TDE emission peak.
However, it could be challenging to distinguish and separate TDEs from other astrophysical sources using the single $g-r$ color.
To address this limitation, a synergistic joint observation with LSST and RST, utilizing multiple-band photometry, is expected to 
enhance TDE search efficiency.
For instance, follow-up RST wide survey observations of LSST-detected TDE candidates can improve high-$z$ TDE selection
(e.g., the F087-F129 color in RST, see Figure~\ref{fig:color_highz_unobs}).
Toward lower redshifts, a few visits with low cadence are adequate for TDE identification.
Detailed quantitative arguments about the photometric selection criteria and observation strategies for detecting $z\sim 2-3$ TDEs are left for future work.

The Euclid Deep survey will cover $\sim 53~{\rm deg}^2$, three times wider than the area of RST Wide.
The expected $5\sigma$ point-source depths are $I_{\rm E} = 28.2$ and $Y_{\rm E}J_{\rm E}H_{\rm E} \simeq 26.4$ mag.
Those coadded depths will be achieved through $\sim 40-50$ visits over six years of operations \citep{Euclid_I_2022,Euclid_XXI_2022}.
i.e., $I_{\rm E} = 25.3$ and $Y_{\rm E}J_{\rm E}H_{\rm E} \simeq 24$ mag for each individual visit \citep[e.g.,][]{Laureijs_2011}.
For the given survey parameters, detecting TDEs at $z\sim 4$ ($Y_{\rm E}\gtrsim 25.5$; see Figure~\ref{fig:SED_highz_unobs}) would 
require $\sim 20$ visits within 30 days after the TDE emission peak. 
However, designing such a high-cadence observation program may pose challenges.

\subsection{Detection numbers of high-$z$ TDEs}
\label{sec:detection}

Finally, we give an estimate of the expected detection number of TDEs from obscured and unobscured AGNs identified in the survey as
\begin{equation}
\mathcal{N}_{\rm TDE} = \mathcal{R}_{{\rm TDE}}(>M_\bullet) ~ \Delta T A_{\rm survey} f_{\rm obs},
\end{equation}
where the observation time is set to $\Delta T=1$ yr, $A_{\rm survey}$ is the survey area, the survey efficiency is defined as
$f_{\rm obs}={\rm min}(\Delta t_{\rm obs}/\Delta T, 1.0)$, and $\Delta t_{\rm obs}$ is the time duration (in the observer frame) when the TDEs are sufficient 
bright for given detection thresholds.
We summarize the detection number of TDEs for each survey program in Table~\ref{Tab:detection}.

The deep JWST imaging surveys, excluding COSMOS-Web, exhibit sufficient sensitivity to detect TDEs occurring within obscured AGNs at $z\sim 4-7$. 
These surveys cover relatively limited areas where abundant populations of BHs with masses $M_\bullet \lesssim 10^6~\msun$ are prevalent, 
as shown in Figure~\ref{fig:tde_rate}.
For such obscured sources, we consider two TDE rates, $\mathcal{R}_{\rm TDE}$, integrated for BH masses $M_\bullet \geq 10^4~\msun$ and $10^5~\msun$. 
Since the TDE emission dominates over the steady AGN emission when $M_\bullet \lesssim 10^5~\msun$, we assess the detectability based on 
the F277W flux density in comparison to the survey's limiting magnitude. 
Additionally, we estimate the observable time window, $\Delta t_{\rm obs}$, from the color-magnitude diagram shown in Figure~\ref{fig:color_highz}.

The COSMOS-Web survey, due to its relatively shallow depth, can only reach the TDE flux densities only when 
the underlying AGN emission significantly contributes to the total SED, requiring BH masses of $M_\bullet \gtrsim 10^6~\msun$. 
However, the observable time window is limited to $\Delta t_{\rm obs}\lesssim 10$ days. 
As a result, the survey efficiency is constrained to $f_{\rm obs}\lesssim 0.03$, as demonstrated in Figures~\ref{fig:SED_z45} and \ref{fig:color_z45}.

The High-latitude Time Domain Survey with RST proves to be sensitive to TDEs occurring within unobscured AGNs at $z\simeq 4-7$.
Both the wide and deep survey tiers cover sufficiently large areas, enabling detection of even rarer unobscured TDEs in AGNs with masses 
$M_\bullet \lesssim 10^6~\msun$.
For these unobscured sources, we consider two TDE rates integrated for BH masses $M_\bullet \geq 10^4~\msun$ and $10^5~\msun$. 
To assess the detectability, we analyze the F129 flux density in relation to the survey's limiting magnitude and calculate the observable time window, $\Delta t_{\rm obs}$, 
from the color-magnitude diagram shown in Figure~\ref{fig:color_highz_unobs}.
Since the depth of the RST surveys can reach the expected TDE flux density through multiple-epoch observations, we can maximize the survey efficiency, 
with $f_{\rm obs}\simeq 1$, for most redshift ranges of $z=4-7$.
Given these survey setups, we predict to detect $N_{\rm TDE}\sim 8~(40)$ TDEs originating from AGNs with $M_\bullet \gtrsim 10^5~(10^4)~\msun$ in one year of operation.
The number of detections decreases with increasing redshift, but remains $N_{\rm TDE}\sim 1~(30)$ even at $z\sim 7$ for the wide-tier survey for the same BH mass range.

It is also worth noting that the COSMOS-Web survey show the capability to detect TDEs in unobscured AGNs at $z\sim 4-7$, 
regardless of the BH mass at the nuclei, as described in Figure~\ref{fig:SED_highz_unobs}.
However, even with the largest survey area planned for JWST surveys, covering $\simeq 0.6$ deg$^2$, the detection is primarily limited to
TDEs occurring in AGNs with the lowest-mass BHs of $M_\bullet \simeq 10^4~\msun$.
To expand the sample size of TDEs detectable in the COSMOS-Web survey, it might be beneficial to consider a transient survey plan that incorporates multiple visits as in 
the RST High-latitude Time Domain Survey. 
Such an observational strategy has the potential to push the boundary for BH detection down to another order of magnitude, 
allowing us to probe the underlying shape of the mass function of seed BHs in a more comprehensive way.

The detection number of TDEs at $z\gtrsim 4$ through a joint observation campaign with LSST and RST depends on the design of the synergy observation programs.
As a showcase, we provide a reference number for the detection, considering a scenario where all TDE candidates selected by LSST ($t_{\rm obs}=60$ days is adopted)
are confirmed as true high-$z$ TDEs through follow-up observations with RST.
The upper bound of detection numbers in this optimistic scenario is $N_{\rm TDE}\sim 51$, $970$, and $5100$ for $z\sim 4$ TDEs
originating from AGNs with BH masses greater than $M_\bullet \gtrsim 10^6$, $10^5$, and $10^4~\msun$, respectively.
A quantitative discussion on the synergy design is beyond the scope of this work and left for future investigation.

\subsection{JWST spectroscopy of high-$z$ TDEs}

JWST spectroscopic follow-up observations of high-$z$ TDEs detected through JWST NIRCam surveys, RST Deep and Wide tier surveys, or 
RST-LSST joint campaigns promise to provide rich insights into these phenomena.
Primarily, the PRISM mode of JWST facilitates the detection of emission lines such as Ly$\alpha$ (for $z>4$) and H$\alpha$ (for $z<6.5$), 
determining the redshift of the source. 
Furthermore, the confirmation of broad-line emission from those TDEs is a definitive signature of massive BHs in their nuclei.
Medium- or high-resolution spectroscopy allows for the estimation of the nuclear BH masses via single-epoch methods 
validated in the low-redshift universe.

Spectral decomposition of narrow and broad components of Balmer lines (e.g., H$\beta$ and H$\alpha$) for unobscured TDEs is feasible  
down to $\Muv \sim -18$ mag for the underlying AGN brightness, particularly when the BH mass exceeds $M_\bullet \gtrsim 0.3-1.0\times 10^6~\msun$ 
(see \citealt{Maiolino_2023_JADES} in the JADES survey).
Such follow-up observations, targeting unobscured TDEs preselected by RST and LSST, can directly constrain the TDE-rate mass function 
(see Figure~\ref{fig:tde_rate}).

For obscured TDEs, deep NIRSpec observations can reach down to $M_\bullet \sim 10^7~\msun$ or lower
(e.g., the dust-reddened AGN, CEERS 746, discussed in \citealt{Kocevski_2023}).
Even without direct BH mass measurements, the TDE rate, primarily dominated by lower-mass BHs, provides constraints on the BHMF beyond $z\gtrsim 4$ for
$10^4\lesssim M_\bullet/\msun \lesssim 10^6$, analogous to discussions on local TDEs ($z\sim 0$) in \citet{Stone_Metzger_2016}.

In cases where broad-line emission is undetected, possibly due to smaller BH masses, the existence of AGNs can still be inferred 
through several methods \citep[e.g., see][]{Maiolino_2023_JADES}.
These include: (1) identifying emission lines with high ionization potential, such as He~{\sc ii} $\lambda 1640$ and $\lambda 4686$ ($E_{\rm ion}=54.4$~eV)
and Ne~{\sc iv} $\lambda 2424$ ($E_{\rm ion}=63.5$~eV), 
(2) observing semi-forbidden nebular lines that indicate high-density gas consistent with a broad line region of an AGN 
(e.g., [N~{\sc iv}] $\lambda 1483$/N~{\sc iv}] $\lambda 1486$ and N~{\sc iii}] $\lambda 1754$/N~{\sc iii}] (total), and
(3) noting broader permitted/semi-forbidden lines in comparison to forbidden lines, as observed in local narrow-line Seyfert 1 galaxies.

\section{Summary and Discussion}\label{sec:discussion}

The unprecedented sensitivity of JWST has enabled the discovery of low-luminosity AGNs at $z\simeq 4-11$, which were hidden in the pre-JWST era.
Spectroscopic follow-up observations have provided estimates of the nuclear BH masses for these sources and pushed the low-mass boundary down to $M_\bullet \sim 10^{6-7}~\msun$. 
Despite this progress, the observed lowest mass of BHs remains $\gtrsim 1-2$ orders of magnitude heavier than the typical mass 
of seed BHs with $\lesssim 10^{6}~\msun$, leaving the mass distribution of early BHs poorly constrained.

In this paper, we focus on UV-to-optical (in rest frame) flares of stellar TDEs embedded in low-luminosity AGNs as a tool to explore low-mass BH populations with $\lesssim 10^{5-6}~\msun$.
Using the AGN host galaxy properties inferred from JWST observations, we estimate the rate of TDEs led by two-body relaxation in a dense stellar cluster.
The TDE rate per galaxy is approximated as $\dot{N}_{\rm TDE} \simeq 2.8\times 10^{-2} \zeta^{5/3}~{\rm yr}^{-1}(M_\bullet/10^6~\msun)^{-1}$, suggesting that lower mass BHs preferentially trigger TDEs.
The parameter $\zeta$ is characterized with the level of dust extinction and size of the nuclear stellar cluster embedded in low-luminosity AGNs (see discussion in Section \ref{sec:cluster}).
For dust-reddened AGNs showing compact morphology reported by JWST observations, we find $\zeta \sim O(1)$, while the value decreases for unobscured AGNs.
The cosmic TDE rate is calculated by convolving $\dot{N}_{\rm TDE}$ with the BHMF model over the redshift range $4\leq z \leq 11$,
which aligns well with the QLF shapes observed in populations at $4\leq z \leq 6$ \citep{Li_2023a, Li_2023b}.
The TDE rate is higher at lower BH masses due to higher event rates per galaxy and greater abundance of lower-mass BHs (see Figures~\ref{fig:MF} and \ref{fig:tde_rate}).

For obscured AGN populations that require deep JWST observations, a substantial number of TDE events are expected within the surveyed area of planned missions.
Conversely, for unobscured AGN populations, the occurrence rate of TDEs is low. 
Therefore, large survey coverage through the RST and GREX-PLUS is needed to identify TDEs in these populations.
The predicted detection numbers for TDEs caused by massive BHs with $M_\bullet \lesssim 10^{4-6}~\msun$ at $z=4-7$ are
$\mathcal{N}_{\rm TDE} \sim 2-10~(0.2-2)$ for the JADES-Medium (and COSMOS-Web) survey with JWST, and 
$\mathcal{N}_{\rm TDE} \sim 2-10~(8-50)$ for the Deep (and Wide) tier of the High-latitude Time Domain Survey with RST
in one-year observations (see Table~\ref{Tab:detection}).
Follow-up observation with RST for TDE candidates pre-selected in the larger LSST sky coverage of $\sim 20,000~{\rm deg}^2$ is expected to significantly
increase the detection number of high-$z$ TDEs.

We conduct SED modeling for high-redshift TDEs embedded in the underlying host AGN continuum and line emission.
Our analysis reveals that the UV-to-optical TDE emission dominates at $\lambda_{\rm rest}\lesssim 5000~{\rm \AA}$ within one year since the emission peak in the observed frame.
Based on the SED modeling, we discuss the selection strategies to detect transient high-redshift TDEs with the ongoing JWST and forthcoming RST observations.
Specifically, we propose color-magnitude selection for high-redshift TDEs in RST transient surveys, utilizing the F087-F129 color and the F129 flux density, 
to distinguish them from low-redshift type Ia SNe with only one or two exposures. 
A long-term ($\sim 1$ yr) observation campaign with RST will enable us to monitor the TDE light curve characterized by a power-law form of $\propto t^{-5/3}$.

This paper focuses on the UV-to-optical emissions originating from high-redshift TDEs. 
A considerable fraction of TDEs may accompany relativistic jets, which can produce prompt gamma-ray and X-ray emissions along with radio/submillimeter afterglow emissions. 
Specifically, in cases involving the disruption of massive OB stars by relatively low-mass BHs, these emissions from jetted TDEs could serve as a viable target for multi-wavelength transient surveys aimed at uncovering high-redshift BH populations~\citep[e.g.,][]{Kashiyama_Inayoshi_2016}.

Furthermore, dense environments such as nuclear star clusters and AGN disks can also serve as potential sources for gravitational waves (GWs) 
induced by the merger of binary BHs
\citep[e.g.,][]{OLeary_2016,Stone_2017a,Bartos_2017,McKernan_2018,Tagawa_2020_agn}. 
Constraints on the low-mass end of the BHMF through high-$z$ TDE detections is crucial to accurately estimate the detection rates of GWs using 
the space-based interferometers such as LISA, Tianqin, and Taiji \citep{Sesana_2008,Luo_2016,Mei_2021,Amaro-Seoane_2023}. 
This not only allows for independent validation of models but also provides insights from a distinct perspective of multi-messenger observations.

\acknowledgments
We greatly thank Xian Chen, Luis C. Ho, Dale D. Kocevski, and Takashi J. Moriya for constructive discussions. 
K.~Inayoshi acknowledges support from the National Natural Science Foundation of China (12073003, 12003003, 11721303, 11991052, 11950410493), 
and the China Manned Space Project (CMS-CSST-2021-A04 and CMS-CSST-2021-A06). 
This work is also supported by Grants-in-Aid for Scientific Research No. JP20K04010, JP20H01904, and JP22H00130 (K.~Kashiyama),
and by Japan Society for the Promotion of Science (JSPS) KAKENHI (20H01939; K.~Ichikawa).

\begin{longrotatetable}
\begin{deluxetable*}{llcccccccccc}
\tablecolumns{1}
\renewcommand\thetable{3} 
\tablewidth{0pt}
\tablecaption{The detection number of TDEs at $z=4-7$}\label{Tab:detection}
\tablehead{
 && &  \multicolumn{4}{c}{Obscured TDEs in 1 yr} && \multicolumn{4}{c}{Unobscured TDEs in 1 yr}  \\
  \cline{4-7} \cline{9-12}
\colhead{Telescope} &\colhead{Survey}&  \colhead{Area} &
\colhead{$N(z\sim 4)$} & \colhead{$N(z\sim 5)$} & \colhead{$N(z\sim 6)$} & \colhead{$N(z\sim 7)$}& &
\colhead{$N(z\sim 4)$} & \colhead{$N(z\sim 5)$} & \colhead{$N(z\sim 6)$} & \colhead{$N(z\sim 7)$}\\
\colhead{(1)}&\colhead{(2)}&\colhead{(3)}&\colhead{(4)}&\colhead{(5)}&\colhead{(6)}&\colhead{(7)}&&
\colhead{(8)}&\colhead{(9)}&\colhead{(10)}&\colhead{(11)}
}
\startdata
JWST&NGDEEP          & 0.0027& 0.26~(1.4) & 0.19~(1.6) & 0.084~(1.1) & 0.03~(0.73) && - & - & -& - \\
          &JADES-Deep    & 0.013  & 1.3~(6.5) & 0.9~(7.8) & 0.4~(5.3) & 0.14~(3.5)    && - & - & - & - \\
          &JADES-Medium & 0.053  & 2.5~(13) & 0.92~(7.8) & 0.4~(5.2) & -   && - & - & - & - \\
          &CEERS             & 0.027  & 0.2~(1.1) & - & - & -   && - & - & - & - \\
          &COSMOS-Web & 0.6      & 0.082 $^\dag$ & - & - & -  && 0.24~(1.2) & 0.17~(1.5) & 0.11~(1.5) & 0.06~(1.4) \\
Roman  &Deep              & 4.20    & - & - & - & - && 1.7~(8.6) & 1.2~(10) & 0.8~(10) & 0.27~(6.5) \\
             &Wide              & 19.04  & - & - & - & - && 7.5~(39) & 5.5~(47) & 3.5~(47) & 1.2~(29) \\
GREX-PLUS   &Deep     & 10  & - & - & - & - && 4.0~(21) & 1.4~(12) & 0.5~(6) & 0.2~(3.7) \\
LSST &&15,000 & &&&&& 51, 970, 5100 $^\ddag$ & - & - & -
\enddata
\tablecomments{~Column (1): telescope. Column (2): survey name. Column (3): survey area in deg$^2$.
Columns (4)-(11): expected detection number of TDEs from obscured and unobscured AGNs at $z\sim 4$, $5$, $6$, and $7$ 
identified in the survey assuming $\Delta z=1$. 
The values indicate the detection numbers for TDEs occurring in AGNs with masses greater than $M_\bullet = 10^5~\msun$ (and $10^4~\msun$).
The observable duty cycle is taken into account (see text for details). For each survey program, we refer to the limiting magnitudes of the F277W filter for JWST,
the F129 filter for RST, and the F232 filter for GREX-PLUS, respectively.
Note that the TDE detection number listed above is calculated by assuming $\bar{m}_\star /\langle m_\star \rangle =4$ for a Salpeter stellar mass distribution
with $0.1 \leq m_\star/\msun \leq 10$, and the value increases as $\propto \bar{m}_\star /\langle m_\star \rangle$ for more top-heavy distribution.\\
$\dag$ For obscured TDE searches in the COSMOS-Web survey, we consider TDEs occurring in AGNs with $M_\bullet \geq 10^6~\msun$;
otherwise the emission in the F277W filter is undetectable.\\
$\ddag$ For the LSST survey, we consider a scenario where {\it all} TDE candidates selected by LSST are confirmed 
as true high-$z$ TDEs through follow-up observations with RST.
The three values are the detection numbers for TDEs triggered by BHs with masses greater than $M_\bullet = 10^6$, $10^5$, and $10^4~\msun$.
The actual number depends on the design of a joint observation campaign with LSST and RST.}
\end{deluxetable*}
\end{longrotatetable}

\bibliography{ref}{}
\bibliographystyle{aasjournal}

\end{document}